\documentclass[longauth]{aa} 
\usepackage{booktabs}
\usepackage{caption}
\usepackage{graphicx}
\usepackage{amsmath}
\usepackage{siunitx}
\usepackage{rotating}
\usepackage{supertabular,booktabs}
\usepackage{soul}
\usepackage{xcolor}
\usepackage{ulem} 

\usepackage{xcolor}
\usepackage[varg]{txfonts} 

\usepackage{natbib,twoopt}
\usepackage[breaklinks=true]{hyperref} 
\hypersetup{ breaklinks=true,colorlinks=true,urlcolor=blue, linkcolor=blue, citecolor=blue, bookmarksopen=true}
\bibpunct{(}{)}{;}{a}{}{,} 

\makeatletter
 \newcommandtwoopt{\citeads}[3][][]{\href{http://adsabs.harvard.edu/abs/#3}%
 {\def\hyper@linkstart##1##2{}%
  \let\hyper@linkend\@empty\citealp[#1][#2]{#3}}}
 \newcommandtwoopt{\citepads}[3][][]{\href{http://adsabs.harvard.edu/abs/#3}%
 {\def\hyper@linkstart##1##2{}%
  \let\hyper@linkend\@empty\citep[#1][#2]{#3}}}
 \newcommandtwoopt{\citetads}[3][][]{\href{http://adsabs.harvard.edu/abs/#3}%
 {\def\hyper@linkstart##1##2{}%
  \let\hyper@linkend\@empty\citet[#1][#2]{#3}}}
 \newcommandtwoopt{\citeyearads}[3][][]%
 {\href{http://adsabs.harvard.edu/abs/#3}
 {\def\hyper@linkstart##1##2{}%
  \let\hyper@linkend\@empty\citeyear[#1][#2]{#3}}}
 \renewcommand*\aa@pageof{, page \thepage{} of \pageref*{LastPage}}
\makeatother

\begin{document}
	
	\title{First MATISSE L-band observations of HD\,179218\thanks{Available at oidb.jmmc.fr}. Is the inner 10~au region rich in carbon dust particles? \thanks{Based on observations collected at the European Southern Observatory, Chile (ESO ID: 0103.D-0069).}}

\author{E.~Kokoulina \inst{\ref{inst_O}} \and
A.~Matter \inst{\ref{inst_O}} \and
B.~Lopez \inst{\ref{inst_O}} \and
E.~Pantin\inst{\ref{inst_Pa}} \and 
N.~Ysard\inst{\ref{inst_Saclay}}\and 
G.~Weigelt \inst{\ref{inst_B}} \and
E.~Habart\inst{\ref{inst_Saclay}}\and
J.~Varga\inst{\ref{inst_L},\ref{inst_K}} \and
A.~Jones \inst{\ref{inst_Saclay}}\and 
A.~Meilland \inst{\ref{inst_O}} \and
E.~Dartois \inst{\ref{inst_Mol}} \and 
L.~Klarmann \inst{\ref{inst_H}} \and
J.-C.~Augereau\inst{\ref{inst_I}} \and
R.~van~Boekel\inst{\ref{inst_H}} \and
M.~Hogerheijde\inst{\ref{inst_L},\ref{inst_P}} \and 
G.~Yoffe \inst{\ref{inst_H}} \and
L.B.F.M.~Waters\inst{\ref{inst_Ra}, \ref{inst_U}} \and
C.~Dominik\inst{\ref{inst_P}} \and 
W.~Jaffe \inst{\ref{inst_L}}\and
F.~Millour \inst{\ref{inst_O}}\and 
Th.~Henning \inst{\ref{inst_H}} \and
K.-H.~Hofmann \inst{\ref{inst_B}} \and
D.~Schertl\inst{\ref{inst_B}}\and 
S.~Lagarde \inst{\ref{inst_O}} \and
R.G.~Petrov \inst{\ref{inst_O}} \and
P.~Antonelli \inst{\ref{inst_O}}\and
F.~Allouche \inst{\ref{inst_O}} \and 
P.~Berio \inst{\ref{inst_O}}  \and 
S.~Robbe-Dubois \inst{\ref{inst_O}} \and
P.~\'Abraham \inst{\ref{inst_K}}\and
U.~Beckmann \inst{\ref{inst_B}} \and
A.~Bensberg \inst{\ref{inst_Ki}} \and 
F.~Bettonvil \inst{\ref{inst_A}} \and
P.~Bristow\inst{\ref{inst_Garch}} \and 
P.~Cruzal\`ebes \inst{\ref{inst_O}} \and
W.C.~Danchi\inst{\ref{inst_Nas}} \and
M.~Dannhoff\inst{\ref{inst_IKP}}\and 
U.~Graser \inst{\ref{inst_H}} 
\and
M.~Heininger \inst{5} \and
L.~Labadie\inst{\ref{inst_C}}\and 
M.~Lehmitz \inst{\ref{inst_H}} \and
C.~Leinert\inst{\ref{inst_H}} \and 
K.~Meisenheimer \inst{\ref{inst_H}} \and
C.~Paladini \inst{\ref{inst_E}}
\and 
I.~Percheron \inst{\ref{inst_Garch}} \and
Ph.~Stee \inst{\ref{inst_O}} \and
J.~Woillez\inst{\ref{inst_Garch}} \and
S.~Wolf\inst{\ref{inst_Ki}} \and 
G.~Zins \inst{\ref{inst_E},\ref{inst_Garch}}
\and
M.~Delbo\inst{\ref{inst_O}}\and 
J.~Drevon \inst{\ref{inst_O}}\and 
J.~Duprat \inst{\ref{inst_IMPMC}}\and
V.~G\'amez Rosas\inst{\ref{inst_L}}\and
V.~Hocdé \inst{\ref{inst_O}}\and
J.~Hron\inst{\ref{inst_V}}\and
C.A.~Hummel\inst{\ref{inst_E}} \and
J.~W.~Isbell\inst{\ref{inst_H}}\and
J.~Leftley \inst{\ref{inst_O}} \and
A.~Soulain\inst{\ref{inst_Sy}}\and
F.~Vakili\inst{\ref{inst_O}}\and
M.~Wittkowski\inst{\ref{inst_Garch}}
}	
	\date{Received <date> / Accepted <date>}

	\abstract
	{Carbon is one of the most abundant components in the Universe. While silicates have been the main focus of solid phase studies in protoplanetary discs (PPDs), little is known about the solid carbon content especially in the planet-forming regions ($\sim $0.1 to 10~au). 
	 Fortunately, several refractory carbonaceous species present C-H bonds (such as hydrogenated nano-diamond and amorphous carbon as well as polycyclic aromatic hydrocarbons (PAHs)), which generate infrared (IR) features that can be used to trace the solid carbon reservoirs. The new mid-IR instrument MATISSE, installed at the Very Large Telescope Interferometer (VLTI), can spatially resolve the inner regions ($\sim$ 1 to 10~au) of PPDs and locate, down to the au-scale, the emission coming from carbon grains.}{Our aim is to provide a consistent view on the radial structure, down to the au-scale, as well as basic physical properties and the nature of the material responsible for the IR continuum emission in the inner disk region around HD\,179218.}{We implemented a temperature-gradient model to interpret the disk IR continuum emission, based on a multiwavelength dataset comprising a broadband spectral energy distribution (SED) and VLTI H-, L-, and N-bands interferometric data obtained in low spectral resolution. Then, we added a ring-like component, representing the carbonaceous L-band features-emitting region, to assess its detectability in future higher spectral resolution observations employing mid-IR interferometry.}{Our temperature-gradient model can consistently reproduce our dataset. We confirmed a spatially extended inner 10~au emission in H- and L-bands, with a homogeneously high temperature ($\sim$1700~K), which we associate with the presence of stochastically heated nano-grains. On the other hand, the N-band emitting region presents a ring-like geometry that starts at about 10~au with a temperature of 400~K. 
	Moreover, the existing low resolution MATISSE data exclude the presence of aromatic carbon grains (i.e., producing the 3.3~$\mathrm{\mu m}$ feature) in close proximity to the star ($\lesssim$ 1~au). Future medium spectral resolution MATISSE data will confirm their presence at larger distances.}{Our best-fit model demonstrates the presence of two separated dust populations: nano-grains that dominate the near- to mid-IR emission in the inner 10~au region and larger grains that dominate the emission outward. The presence of such nano-grains in the highly irradiated inner 10~au region of HD\,179218 requires a replenishment process.
	Considering the expected lifetime of carbon nano-grains from The Heterogeneous dust Evolution Model for Interstellar Solids (THEMIS model), the estimated disk accretion inflow of HD\,179218 could significantly contribute to feed the inner 10~au region in nano-grains. Moreover, we also expect a local regeneration of those nano-grains by the photo-fragmentation of larger aggregates.}
	
	
	\maketitle
	
	\section{Introduction}
The great diversity of exoplanetary systems detected so far \footnote{\href{https://exoplanetarchive.ipac.caltech.edu/}{https://exoplanetarchive.ipac.caltech.edu/}} has raised numerous questions about the composition of planets.
protoplanetary disks (PPDs) are the environments in which planets are formed and dust grains are their building blocks.
In particular, tracing the composition and properties of solid materials in the inner disk regions ($\sim$0.1 to 10~au) is required to unveil the conditions of formation of telluric planets.

 Ubiquitously detected in PPDs \citep[e.g.,][]{2001Bouwman, 2010Juhasz} and in the interstellar medium \citep[ISM;][]{2003Draine}, silicate minerals have been one main focus of solid phase studies because of their strong infrared (IR) signatures around 10 and 20~$\mathrm{\mu m}$. However, carbon-based species, both in gas and solid form, also play a critical role in PPD evolution and planet formation. 
 Carbon is a key parameter in the final composition of planets and of their atmosphere 
 \citep[e.g.,][]{2011Oberg}. Notably, \citet{2018Eistrup} linked the observed atmospheric C/O ratios in exoplanets to formation locations and specific conditions in the disk midplane.
Moreover, carbonaceous ices are essential for forming complex molecules such as hydrocarbons \citep{2018Eistrup}. According to \citet{2010Lee_Nomura}, icy bodies (in the outer disk regions) have a high abundance of carbon in the form of ices and amorphous carbon grains.

While detected in the form of carbonaceous ices, for example, CO in the outer disk regions \citep{2014Pontoppidan}, little is known about the solid carbon species in the inner disk regions and key questions are whether there is a significant reservoir of C-based solid species in the inner 10~au region of disks, in what form are those C-based species and how can solid carbon be incorporated into planetesimals and ultimately planets.
These questions are relevant to understanding the apparent carbon deficit in the primitive inner proto-solar nebula \citep[e.g.,][]{2010Lee_Nomura}. Possible explanations for such a carbon deficit include carbon burning \citep{2010Lee_Nomura} or its removal due to solar radiation pressure \citep{2002Koehler}.


Spectrally featureless or low band contrast species, like non-hydrogenated amorphous carbon and graphite, cannot be efficiently employed to trace the carbon content of disks. This is also the case for the volatile carbonaceous compounds (e.g., ${\rm CO, CO_2, CH_4}$), which cannot be preserved in solid form in the innermost disk regions. Unlike those components, carbonaceous nano-grains presenting C-H bonds (e.g., hydrogenated nanodiamonds, hydrogenated amorphous carbons, PAHs) can be used to trace the solid carbon reservoirs.
Indeed, these species exhibit narrow emission bands in L- and N-band from IR fluorescence following stochastic heating events by UV photons \citep[e.g.,][]{2001Draine, 1984Leger}. 
Amorphous carbonaceous materials are expected to dominate the ISM mass budget of carbon dust \citep[e.g.,][]{2007DartoisCaro, 2004Zubko, 2001Li, 2013Jones}, with a significant part in the form of nano-grains. For instance, in the case of the heterogeneous dust evolution model for interstellar solids (THEMIS)\footnote{\href{https://www.ias.u-psud.fr/themis}{https://www.ias.u-psud.fr/themis}} grain model for the diffuse ISM, the smallest nanograins (size $<$ 0.7 nm) represent about 30\% of the total solid carbon mass.

Carbonaceous nano-grains were first discovered in the interstellar medium as unidentified IR features by \citet{1973Gillett} and later associated with PAHs \citep{1984Leger} and hydrogenated amorphous carbons \citep{1980Wickramasinghe}. Aromatic infrared emission bands are commonly ($\sim$ 70\%) detected around Herbig Ae stars \citep[e.g.,][]{2010Acke} and more sparsely ($\sim$ 10\%) around T Tauri stars \citep{2007Geers}. 
In the very low density ISM, the lifetime of carbonaceous dust is shorter than that of silicate dust \citep[e.g.,][]{2008Serra,2011Jones} but little is known in other environments. In particular, the inner regions of PPDs differ greatly from the ISM in terms of irradiation conditions and densities. 
Up to now, the spatial distribution of carbon species has mainly been performed at large scales ($> 10$ au)
using mid-infrared spectro-imaging instruments such as VLT/NaCo, VLT/VISIR or GTC/CanariCam \citep{2004Habart, 2006Habart, 2007Geers, 2013Maaskant,2018Anas, 2006Lagage}. 
Interestingly, the morphology of images in the carbon features and micron-sized dust emission suggests that the inner regions of some PPDs have been cleared from micron-sized dust but not from carbonaceous nano-particles \citep[e.g.,][]{2007Geers,2013Maaskant}.
In that context, accurate constraints are missing on the spatial distribution and properties of carbonaceous dust particles down to the astronomical unit in the inner disk regions. That is yet essential to determine whether or not the inner regions of PPDs are inherently depleted in solid carbon and to understand the physics, radiative properties and chemistry of these grains.


 The VLTI instrument MATISSE \citep{2014Msngr.157....5L} offers a unique opportunity to study the spatial distribution and characteristics of C-based nano-grains. MATISSE is, so far, the only spectro-interferometer that can access both the L (2.8--4.2~$\mathrm{\mu m}$), M (4.5--5~$\mathrm{\mu m}$) and N (8--13~$\mathrm{\mu m}$) bands. It combines four telescopes of the VLTI, with baseline lengths from about 10 to 150~m, thus providing up to milliarcsecond angular resolution scales. That represents a fraction of the astronomical unit for Young Stellar Objects (YSOs) located at distances less than 150~pc. Moreover, MATISSE provides different spectral resolutions ($R\sim$ 30, 500, 900, and 3400) to study and characterize the diversity of IR spectral features of carbonaceous species.


In this article, we report on the first L-band interferometric observations of the Herbig star HD\,179218 (alias MWC\,614). It is a luminous star ($\sim 100~ \mathrm{L_{\odot}}$) of A0/B9 spectral type with a temperature of $T_{\rm eff}= 9500^{+200}_{-200} ~K$ and stellar radius of $R_*=2.8 ^{+0.9}_{-0.9}~\mathrm{R_{\odot}}$ \citep{2017Mendigutia}, located at a distance of $266^{+5}_{-5}$~pc \citep{2018Vioque}. 
It is surrounded by a pretransitional disk with a ring-like N-band continuum emission from $\mu$m-sized grains located at about 10~au from the central star \citep{2008Fedele}. It was identified as a group-Ia disk source following the classification from \citet{2001Meeus}, which suggests a flared disk structure. Based on spatially unresolved mid-infrared spectroscopy, the circumstellar silicate dust component seems to be highly processed and crystalline \citep{2010Juhasz}, with a significant contribution from enstatite. The latter would be produced in the inner regions and transported to larger distances by radial mixing \citep{2005vanBoekel}.
More recently, from continuum near and mid-infrared interferometric observations in H-, K- and N-bands, \citet{2018Kluska} spatially resolved the NIR-emitting region, which extends over the inner 10~au region. Such an extended NIR-emitting region was confirmed by \cite{2019Perraut} from K-band VLTI/GRAVITY observations 
and could be due to stochastically heated small particles like carbonaceous nano-grains \citep{2017Klarmann} filling the inner disk region. Numerous spectroscopic studies (e.g., \citet{2004Acke_Ancker}, \citet{2010Acke} and \citet{2017Seok}) revealed carbonaceous emission features at 3.3, 6.2, 7.7, 8.6, 11.3, and 12.7~$\mathrm{\mu m}$ for that object. In addition, more recent studies provided first constraints on the spatial distribution of those carbonaceous emission features. In N-band, \citet{2018Anas} spatially resolved the two most prominent aromatic bands at 8.6 and 11.3~$\mathrm{\mu m}$ and suggested that the emission at 8.6 and 11.3~$\mathrm{\mu m}$ extends out to larger radii than the N-band dust continuum emission. 
\citet{2019Bouteraon} spatially resolved the $\sim$ 3.3~$\mathrm{\mu m}$ and $\sim$3.4~$\mathrm{\mu m}$ features down to a distance of 30~au from the central star. Those two features originate from C-H bonds associated with aromatic rings and aliphatic chains, respectively. They also showed that the 3.3/3.4~$\mathrm{\mu m}$ band ratio does not change significantly with the distance from the star. Since aliphatic grains are expected to be more fragile against UV processing (e.g., \citet{2010Acke}), that suggests a replenishment and a regeneration of those grains to maintain the strength of the 3.4~$\mathrm{\mu m}$ band, relative to the 3.3~$\mathrm{\mu m}$ band, as we approach the star.


Based on a multiwavelength dataset, the aim of our study is to provide a consistent view on the radial structure, down to the au-scale, and basic physical properties of the material responsible for the IR continuum emission in the inner disk region around HD\,179218. 
Moreover, we aim to provide first constraints on the spatial distribution of aromatic carbon nano-grains, through the 3.3~$\mathrm{\mu m}$ feature, in the inner 1-10~au region. Finally, in the perspective of future MATISSE observations in medium spectral resolution ($R\sim500$), synthetic L-band visibilities are produced to assess further the feasibility of detection of carbonaceous nano-grains in the 1-10~au region.

The paper is organized as follows: Section \ref{sec:observations} summarizes the new observations with VLTI/MATISSE which complement the existing data obtained with two other VLTI instruments, MIDI and PIONIER. Section~\ref{sec:continuum} describes a geometrical modeling of the MATISSE data, followed by a temperature-gradient modeling of our multiwavelength dataset.
In Section~\ref{sec: PAHs} we investigate the presence of the carbonaceous features in the obtained low spectral resolution MATISSE data and provide the limits on the detection of carbonaceous nano-grains based on synthetic medium spectral resolution MATISSE data.  Section~\ref{sec:discussion} provides an interpretation of our modeling results based on the physics of nano-grains. 
Section~\ref{Conclusions} summarizes our work and outlines the perspectives with upcoming MATISSE observations in higher spectral resolution. 


 \section{Observations} \label{sec:observations}
 \subsection{MATISSE observations}


One snapshot of HD\,179218 in low spectral resolution in L-band was observed with MATISSE on March 24, 2019 with the small AT quadruplet giving 6 simultaneous $uv$ points. The corresponding baseline lengths ranges from 7 to 25~m. The N-band flux of HD\,179218 ($\sim 10$~Jy) is lower than the sensitivity limit of MATISSE with ATs for absolute visibility measurements ($\sim 15$~Jy), therefore no N-band photometry was acquired. Since chopping is only performed during that step, no L-band data with chopping was acquired either. As a consequence, the L-band thermal background emission was removed using sky measurements done at the beginning of the observation (a few minutes apart).
Table \ref{tab:obs. log} gives an overview on the observations. 

 We reduced the data with the version (1.5.0) of the standard MATISSE data reduction pipeline. The data reduction steps of MATISSE are described in \citet{Millour2016}. We obtained six dispersed L-band absolute visibilities, three independent closure phases (out of four closure phases in total) and four total flux measurements. 
 In the study presented here, we aim to measure the dust IR continuum emission and to assess the limits of the detection of the carbonaceous features. A detailed study of the brightness asymmetries suggested by the MATISSE closure phases will be carried out in a future study. For now, we refer to Appendix~\ref{sec:closure_phase} for a first modeling of the MATISSE closure phases performed in the frame of our temperature-gradient model. 
 Considering our primary goal of this study, we thus focused on the visibilities and total flux for which two calibration procedures were applied: 
\begin{itemize}
 \item The calibrated visibilities were obtained by dividing each raw visibility measurement of HD\,179218 by the instrumental visibility measured on the associated interferometric calibrator observed just after. 
 \item The calibrated total photometric fluxes were obtained by multiplying the ratio of the target-to-calibration-star raw fluxes, measured by MATISSE at each wavelength, followed by multiplication by the absolute flux of the corresponding calibrator. 
\end{itemize}
Both routines were performed using python tools of the MATISSE consortium. 
The interferometric and spectrophotometric calibrator was selected from Mid-infrared stellar Diameters and Fluxes compilation catalog (MDFC), a calibrator catalog for mid-infrared interferometric observations \citep{2019MNRAS.490.3158C}.


In Fig.~\ref{fig:all models} in light blue curves visibilities are shown as a function of wavelength for different baselines. Since the MATISSE L-band visibility at the shortest baseline ($B_{\rm proj} \sim 7$~m) is  already rather low (around 0.4), and even lower for larger baselines, it suggests a spatially extended L-band emitting region. At $B_{\rm proj} \sim 7$~m, that translates to a spatial resolution of $\lambda/2B_{\rm proj} \sim 14$~au at the 266~pc distance of HD\,179218. 
 Then, the visibilities at higher spatial frequency seem to reach a plateau around 0.15, which likely represents the flux contribution from the unresolved star with respect to the total emission, with a circumstellar emission that is fully resolved by MATISSE (see Fig.~\ref{Fig: all temperature gradient fits} upper left corner). The increasing profile (with spatial frequency) of the group of five visibilities (at high spatial frequencies) is likely due to the chromaticity of the stellar-to-total flux ratio over the L-band.

The L-band calibrated flux is represented in green on the right hand side of Fig.~\ref{fig:SEDs}. 
The aromatic feature at 3.3~$\mathrm{\mu m}$ seems to be present in the calibrated MATISSE L-band spectrum. However, it is possible that such feature is due to several telluric absorption lines that are mixed together due to the low spectral resolution. If we have had two nearby calibrators, which is not the case for our set of data, we could have evaluated the amplitude of that effect. We further analyze the MATISSE L-band spectrum, on the basis of a carbon nano-grain emission template, in Section~\ref{sec: PAHs}.


\subsection{MIDI data}
The calibrated MIDI N-band visibilities from \citet{2015Menu} were taken from the Optical interferometry Database \citep[OiDB][]{2014SPIE.9146E..0OH} of the Jean-Marie Mariotti Center \footnote{Available at oidb.jmmc.fr} (see Appendix~\ref{app:data log} for the MIDI observation log). Decommissioned in 2014, the
MIDI instrument \citep{2003Leinert} combined the light from two telescopes of the VLTI in the
mid-infrared (MIR) (8--13~$\mathrm{\mu m}$) with two
spectral resolutions ($R$ = 30 and 230).
The HD\,179218 low spectral resolution MIDI data consist of 27 observations with baseline lengths ranging from 10 - 90~m (see Appendix \ref{app:data log}).


As shown in Fig.~\ref{Fig: all temperature gradient fits} (upper right corner), the MIDI visibilities show characteristic sinusoidal modulations at
the longest baselines ($\sim$50 to 85~m), which are reminiscent of an object having a brightness distribution with sharp edges like a
uniform disc, ring or a binary system with two unresolved components. According to \citet{2008Fedele}, the latter can be excluded. 

\subsection{PIONIER data}
The calibrated PIONIER \citep{2011Bouquin} H-band visibilities were also taken from OIDB.
The PIONIER interferometric observations (see Appendix~\ref{app:data log} for the PIONIER observation log) were carried out using the 1.8~m auxiliary telescopes (AT) of the VLTI with a four-
telescope recombiner operating in H-band (1.55--1.80~$\mathrm{\mu m}$). The three AT configurations provide different baselines in the range 10--140~m. In Fig.~\ref{Fig: all temperature gradient fits} at the bottom left, we see a series of inclined lines. Each line corresponds to a $uv$ point, which is dispersed in the six wavelenght channels. The top of each line corresponds to the short edge of the H-band while the bottom of each line corresponds to the long edge of the H-band. 
The dispersed H-band visibilities are rather low and do not vary with spatial frequency. \citet{2018Kluska} mentioned that the circumstellar flux is over-resolved because the visibility reaches a plateau before our shortest baseline. 
The value of that visibility plateau is set by the flux ratio star/(star+disc), which depends strongly on the wavelength (stronger at short wavelengths than at longer wavelengths). 

\subsection{Photometric observations}
The visible and IR broadband SED used in our temperature-gradient modeling is shown in Fig.~\ref{fig:SEDs} (dotted black line) and interpolated measurements in Fig.~\ref{Fig: all temperature gradient fits} (bottom right panel, dotted blue line). The SED is based on various photometric measurements taken from the VIZIER\footnote{\href{http://vizier.u-strasbg.fr/vizier/sed/}{http://vizier.u-strasbg.fr/vizier/sed/}} database. 

For most of the photometric measurements, the error bars are $\sim$ 5--6\%. We thus assumed a typical 5\% error for all the interpolated data points. For sake of comparison, we also considered the high-resolution MIR Spitzer spectrum, taken from the Spitzer archive\footnote{\href{https://irsa.ipac.caltech.edu/data/SPITZER/docs/spitzerdataarchives/}{https://irsa.ipac.caltech.edu/data/SPITZER/docs/spitzerdataarchives/}} and SWS spectrum from the ISO archive\footnote{\href{https://irsa.ipac.caltech.edu/data/SWS/}{https://irsa.ipac.caltech.edu/data/SWS/}}. The ISO SWS data presented here is before normalization of each individual spectral segments and with the regions of overlap intact. The observed SED was de-reddened using the visual extinction $A_V = 0.53$ \citep{2018Vioque}.

	\begin{table*}
	\caption{Overview of the MATISSE observations of HD\,179218. $\tau_0 $ is the atmospheric visible coherence time. LDD is the estimated angular diameter of the calibrator. 
		}
		\begin{center}
			\label{tab:obs. log}
			\begin{tabular}{c c c c c c c c c}
				\hline
				\hline
				Date and time (UTC)& Seeing & $\tau_0$ & Stations & Configuration & Science and Calibrator & LDD \\ 
				 & ($''$) & (ms) & & & & (mas) \\
				\hline
				2019-03-24 09:28 & 0.61 & 7.3 & A0-B2-D0-C1&small&HD\,179218 &\\
				
				2019-03-24 09:50	& 0.72 & 6.9 &A0-B2-D0-C1	& small & $\gamma$ Aql & 7.26 \\

			
				\hline
			\end{tabular}
		\end{center}
	\end{table*}

 \section{IR continuum modeling}\label{sec:continuum}
In this section, we describe our modeling of the IR continuum emission of HD\,179218. 
Between the H-band and N-band explored by \citet{2018Kluska}, we expect to probe 
the inner 10~au region together with 
the ring-like geometry localized around 10~au.
Hence, two modeling steps were performed: 1) a geometrical model of the MATISSE L-band visibilities to derive the fundamental characteristics of the L-band emitting region, and 2) a multiwavelength temperature-gradient model combining the IR broadband SED, the H-band PIONIER visibilities, the L-band MATISSE visibilities, and the N-band MIDI visibilities, to obtain a global description of the inner continuum emission of the HD\,179218's disc.




\subsection{Geometrical approach}\label{sec:geom.model}
\subsubsection{Model descriptions}
In order to characterize the morphology of the object as constrained by the observed MATISSE visibilities in L-band, we decided to use a similar geometrical approach as in \citet{2018Kluska}. The NIR and MIR data were best reproduced with two different simple geometries for the circumstellar emission: 1) a 1D Gaussian brightness distribution with a FWHM of $\sim$ 15~au in the NIR.
2) a ring of radius 12.4~au (and convolved with a Gaussian with FWHM of 5.4~au) in the MIR.


Following Kluska's approach, MATISSE L-band data can be represented by either of the geometries. In our geometrical modeling approach, we thus consider two models:
\begin{itemize}
 \item a Gaussian,
 \item a ring convolved with a Gaussian,
\end{itemize}

The model visibility is calculated by the ratio of the correlated flux to the total flux :
\begin{equation}
 V_\mathrm{tot}(B/\lambda) = \frac{|F_\mathrm{corr, ~star}(B/\lambda) + F_\mathrm{corr, ~disc}(B/\lambda)|}{F_\mathrm{star}(\lambda) + F_\mathrm{disc}(\lambda)}
 \label{eq:total visib}
\end{equation}
with $F_\mathrm{corr, ~star}(B/\lambda)$ and $F_\mathrm{corr, ~disc}(B/\lambda)$ the correlated flux of the star and the circumstellar emission, respectively, and $F_\mathrm{star}(\lambda)$ and $F_\mathrm{disc}(\lambda)$ their total fluxes. 
The correlated flux represents the modulus of the Fourier Transform of the object brightness distribution at the spatial frequency $B/\lambda$, with $B$ the baseline length. Equation~\ref{eq:total visib} can be rewritten as:
\begin{equation}\label{eq: total vis with star visib}
 V_{tot}(B_p/\lambda) = \frac{|F_\mathrm{star}(\lambda) \cdot V_\mathrm{star}(\frac{B_p}{\lambda})+ F_\mathrm{disc}(\lambda) \cdot V_\mathrm{disc}(\frac{B_p(i, \theta)}{\lambda})|}{F_\mathrm{star}(\lambda) + F_\mathrm{disc}(\lambda)}
\end{equation}
with $V_\mathrm{star}(B_p/\lambda)$ and $V_\mathrm{disc}(B_p(i, \theta)/\lambda)$ as the corresponding visibilities. $B_p=B_{u}^2 + B_{v}^2$ is the projected baseline length with $B_{u}$ and $B_{v}$ the coordinates of the baseline vector projected on the sky plane.
The star is considered to be unresolved by MATISSE with the current VLTI baseline lengths, that is $V_\mathrm{star}(B_p/\lambda) = 1$. 
$B_p(i, \theta) = \sqrt{B_{u,\theta}^2 + B_{v, \theta}^2 \cos^2(i)}$ is the length of the projected baseline expressed in the reference frame rotated by an angle $\theta$, with:
\begin{equation}
 B_{u, \theta} = B_{u}\cos \theta + B_{v} \sin \theta
 \end{equation}
 \begin{equation*}
 B_{v, \theta}=-B_{u} \sin\theta +B_v \cos \theta. 
 \end{equation*}
All models are thus rotated and compensated for the flattening introduced by the inclination $i$ and position angle $\theta$ considered for the circumstellar emission.


\paragraph{{\bf Gaussian model}}
\label{Section:first geometrical}
 The circumstellar emission is represented by a Gaussian brightness distribution characterized by a FWHM $\omega_g$, an inclination $i_{g}$ and a position angle $\theta_G$.
 The corresponding correlated flux is:
\begin{multline}
F_\mathrm{corr,~G}\left(B_p(i_G, \theta_G)/\lambda \right)=\\ = \pi \left(\frac{\omega_G}{2\sqrt{\ln{2}}}\right)^2 \exp{\left(-\left(\frac{\pi\omega_G}{2\sqrt{\ln{2}}}\frac{B_p(i_G, \theta_G)}{\lambda}\right)^2\right)}B_{\lambda}(T_G)
\end{multline}
 
where $\pi \left(\frac{\omega_G}{2\sqrt{\ln{2}}}\right)^2$ is the solid angle subtended by the Gaussian brightness distribution and $B_{\lambda}(T)$ is the Planck function.
 
 
The total visibility of the star + Gaussian model thus becomes:
\begin{equation}
 V_\mathrm{tot}(B_p(i_G, \theta_G)/\lambda) = \frac{\mid F_\mathrm{star}(\lambda) + F_{corr,~ G}\left(B_p(i_G, \theta_G)/\lambda \right)\mid}{F_\mathrm{star}(\lambda) + F_{G}(\lambda)}
\end{equation}
 with $F_\mathrm{star}(\lambda) = \pi \left( \frac{R_{\star}}{d}\right)^2 B_{\lambda}(T_{\star})$ is a total stellar flux and $F_{G}(\lambda)$ is the total flux of the Gaussian circumstellar emission. $R_{\star}$ and $T_{\star}$ are the radius and the effective temperature of the star respectively. 
\paragraph{{\bf Ring model}}
The circumstellar brightness distribution is here represented by a ring convolved with a Gaussian. The corresponding correlated flux is equal to:
\begin{multline}
 F_\mathrm{corr, ~r \ast G}\left(B_p(i_{r \ast G}, \theta)/\lambda \right) = \pi\left(\frac{\omega_{r \ast G}}{2\sqrt{\ln{2}}}\right)^2 J_0 \left( 2 \pi r \frac{B_p\left(i_{r \ast G}, \theta_{r \ast G}\right)}{\lambda} \right) \\ \cdot \exp{\left(-\left(\frac{\pi\omega_{r \ast G}}{2\sqrt{\ln{2}}}\frac{B_p(i_{r \ast G}, \theta_{r \ast G})}{\lambda}\right)^2\right)}B_{\lambda}(T_{r \ast G})
\end{multline}
\
where 
$i_{r \ast G}$, $\theta_{r \ast G}$ are the inclination and position angle of the projected baselines, $r$ is an angular radius of the ring, and $\omega_{r \ast G}$ is the FWHM of the Gaussian. $J_0$ is the zeroth-order Bessel function of the first kind, $B_p(i_{r \ast G}, \theta_{r \ast G})$ 
is the baseline length as defined in Section~\ref{Section:first geometrical}. 
The total visibility is given by: 
\begin{equation}
 V_\mathrm{tot}\left(B_p(i_{r \ast G}, \theta_{r \ast G})/\lambda \right) = \frac{\mid F_\mathrm{star}(\lambda) + F_\mathrm{corr,~r \ast G }\left(B_p(i_{r \ast G})/\lambda \right)\mid}{F_\mathrm{star}(\lambda) + F_{r \ast G}(\lambda)},
\end{equation}
with $F_{r \ast G}(\lambda)$ as the total flux of the ring component.
\subsubsection{Fitting approach}\label{subsec:Fitting approach}


Our fitting approach is based on the Bayesian inference method \emph{emcee} \citep{2013EMCEE}, a python-based implementation of the Markov Chain Monte-Carlo (MCMC) algorithm. It allows to obtain the minimum $\chi^2$ value over the considered multidimensional parameter space. Moreover, it provides the marginal probability density function associated with each free parameter and their associated quantiles like the median. 
Among the different model parameters, we fix the inclination and position angle of the circumstellar components. For the one-component model, they were fixed to the values derived by \citet{2018Kluska}, namely: $i_G$ = 54.8$^{\circ}$ and $\theta_G$ = 25.5$^{\circ}$ (Gaussian model), and $i_{r \ast G}$ = 52.5$^{\circ}$ and $\theta_{r \ast G}$ = 26.4$^{\circ}$ (Ring model). 
The list of free parameters and their associated priors are given in Table \ref{table:1 disc models} for the one-component models.
In the following, the best-fit values of the free parameters will correspond to the values associated with the least $\chi^2$ value. In addition, the first and third quantiles of the marginal probability distributions, around the median, will also be provided as an estimate of the probability distribution dispersion and thus of the parameter error bars. 

\subsubsection{Modeling results}\label{subsec:geom_mod_results}
 Table \ref{table:1 disc models} shows, for the one-component models, the best-fit values, along with the median values with $\pm$1 $\sigma$ (corresponding to the first and third quantiles).

\begin{figure*}
 \centering
 \includegraphics[width=17cm]{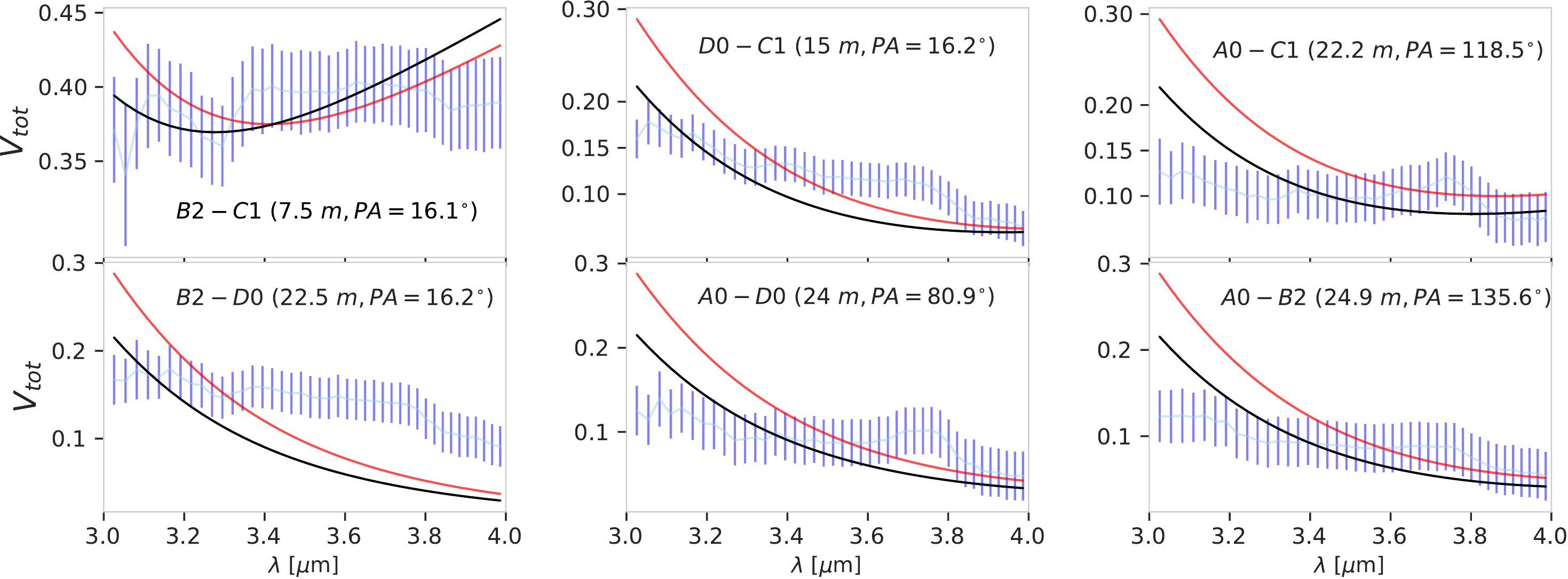}
 \caption{Visibility as a function of wavelength for different baselines with geometric best-fit models (see Section~\ref{sec:geom.model}): Gaussian (red line) and a ring convolved with a Gaussian (black line). The telescope pairs, projected baseline length ($B$) and the position angles of the projected baseline ($PA$) are shown in each subplot.}
 \label{fig:all models}
\end{figure*}

In the case of the one-component Gaussian and ring models, the MATISSE visibilities favored a Gaussian distribution with a FWHM of about 10.65~au and a ring of 3.17~au radius convolved by a Gaussian with a FWHM of $\sim$ 13.34~au, respectively. The Gaussian model leads to a better $\chi^2$ than the ring model and is in agreement with the value derived in the NIR by \citet{2018Kluska}. We note that the best-fit ring model approaches the aspect of a Gaussian given the large FWHM of the convolving Gaussian compared to the ring radius. As shown in Fig.~\ref{fig:all models} (red and black lines), the corresponding best-fit L-band visibilities reproduce the level of the measured visibilities, especially for the shortest baseline. However, we notice larger discrepancies for the larger baseline visibilities, especially at short wavelengths, which illustrates the limits of those simple one-component models. Overall, with this first modeling step, it seems that the MATISSE visibilities favor an extended inner emission similar to what is seen in the NIR.


 Finally, the Gaussian and ring models present a similar overall temperature of 431 K and 448 K, respectively. Those temperatures are a little bit higher than the expected equilibrium temperature of a gray grain that would be located at the half width at half maximum of the Gaussian model, i.e., $\sim$ 5~au from the star. Indeed, the temperature of a gray grain in radiative equilibrium with the stellar emission, and located at a distance $r$ from the star, can be approximated by:
\begin{equation} \label{eq:equil_temp} 
T_{gray} = T_{\star}\cdot \sqrt{R_{\star}/2r} \end{equation}
At a distance of 5~au from the star, the expected equilibrium temperature would be 390~K. The difference between 431--448~K and 390~K is not meaningful and due to our model approximations.

\begin{table*}\centering
\captionof{table}{Free parameters with their priors for the one component model. The best-fit values, with the associated reduced $\chi^2$, and the median values, with the associated $1\sigma$ error bars, are indicated.}
\begin{tabular}{ c c c c c c c}
\hline\hline
& \multicolumn{4}{c}{One-component Models} \\
\hline
&& & \multicolumn{2}{c}{Gaussian} & \multicolumn{2}{c}{Ring $\ast$ Gaussian} \\
reduced $\chi^2$ & & & \multicolumn{2}{c}{1.33} & \multicolumn{2}{c}{2.65} \\
\hline
 Description &Parameter & Prior & Median & Best-fit & Median & Best-fit\\

Radius&$r~[\mathrm{au}]$ & from 0 to 35 & - & - & $4.42^{+2.18}_{-2.82}$ &3.17 \\
 FWHM & $\omega_G~ [\mathrm{au}]$ & from 0 to 35 & $10.65^{+1.42}_{-1.17}$ & 10.65 & - & -\\
 FWHM & $\omega_{r \ast G}~ [\mathrm{au}]$ & from 0 to 35 & - & - &$12.11^{+2.65}_{-5.59}$ & 13.34\\
Temperature & $T_G~[\mathrm{K}]$  & from 100 to 1000 & $437.56^{+34.27}_{-21.26}$ &431.65& -&-\\
Temperature & $T_{r \ast G}~[\mathrm{K}]$  & from 100 to 1000 & - &-& $452.48^{+24}_{-22.24}$ &448\\
\hline
\end{tabular}
\label{table:1 disc models}
\end{table*}

Overall, it seems that no further meaningful constraint can be put on the L-band emitting region from these simple one-component geometrical models. Hereafter, we thus describe the use of a temperature-gradient model to provide a global view on the geometry of the IR continuum emission.

 \subsection{Temperature-gradient approach}\label{subsec:temp_grad model}
\subsubsection{Model description}
To model more accurately the radial structure of the disc, we applied a temperature-gradient model to a multiwavelength dataset that includes NIR and MIR interferometric data and an IR broadband SED. With such a temperature-gradient (and surface-density) model, we can take into account the fact that different radial disk regions will emit at different wavelengths, with different temperatures, and different optical depths along the line of sight. Such model aims to reproduce the emission from the disk surface as seen in the IR.
Here, the circumstellar emission is produced by a geometrically flat temperature-gradient dusty disk that has temperature and surface density radial profiles given by:
\begin{equation} \label{eq:temp} 
 T_r =T_{in}\Big(\frac{r}{r_{in}}\Big)^{-q},
 \end{equation}
 \begin{equation}
  \Sigma_{r} =\Sigma_{in}\Big(\frac{r}{r_{in}}\Big)^{-p},
  \label{eq:surf density}
 \end{equation}
 where $r_{in}$ is the inner radius, $T_{in}$-temperature at the inner radius and $\Sigma_{in}$ is the dust grain surface density in $\mathrm{g/cm^2}$ at the inner radius respectively; and $p$ and $q$, the surface density and temperature power-law exponents, respectively. 
 Such a temperature-gradient model splits the disk into infinitesimal ringlets emitting like blackbodies. Each ringlet is located at a distance $r$ from the star and its
blackbody emission is weighted by an emissivity factor, dependent on $\lambda$ (but that we assume in our modeling process to be achromatic),
\begin{equation}\label{eq:emissivity factor}
 \epsilon_{\tau} = 1 - e^{(-\tau_{r, \lambda}/\cos{i})} 
\end{equation}
where $\tau_{r, \lambda}$ is the vertical optical depth and is the product of the surface density $\Sigma_{(r)}$ and mass absorption cross section $\kappa_{\lambda}$: 
\begin{equation}
\tau_{r, \lambda} = \kappa_{\lambda}\Sigma_{r}.
\end{equation}
The disk mass and surface density are linked by the following expression:
\begin{equation}
M_{dust} =\int_{r_{in}}^{r_{out}} 2 \pi r \Sigma_{r}dr,
\label{eq:disc mass}
\end{equation}
using Eq. \ref{eq:surf density} and \ref{eq:disc mass}, we find the relation between $M_{dust}$ and $\Sigma_{in}$:
\begin{equation}
\Sigma_{in} = \frac{M_{dust}}{2 \pi r_{in}^2 f},
\end{equation}
with $f$:
\begin{equation}
f = \frac{1}{2 - p}\Big[\Big(\frac{r_{out}}{r_{in}}\Big)^{(2-p)}- 1\Big].
\end{equation}
The disk flux density is given by:
\begin{equation}\label{eq:1}
F_{\lambda,~ disc}(i) = \frac{2 \pi \cos i}{d^2} \int_{r_{in	}}^{r_{out}} B_{\lambda}(T_r)\varepsilon_{\tau} rdr,
\end{equation}
and its visibility is defined as in \citet{2014Matter}: 
\begin{equation}
V_{\lambda,~disc}(B_p(i,\theta)) = \frac{2\pi}{d^2} \frac{1}{F_{\lambda}(0)} \int_{r_{in	}}^{r_{out}} B_{\lambda}(T_r)\varepsilon_{\tau}J_0\Big[\frac{2\pi}{\lambda}B_p(i, \theta)\frac{r}{d}\Big]rdr,
\end{equation}


$B_{\lambda}(T_r)$ the Planck function, and $d$ the distance to the object. 
From that temperature-gradient description, 
we built a two-component disk model with the aim to reproduce, in a consistent way, the complete dataset comprising the IR broadband SED, the PIONIER H-band, the MATISSE L- and the MIDI N-band visibilities. 

\subsubsection{Fitting approach} \label{subsec:fitting}
We included two components in our temperature-gradient model. First, an inner component that we considered to be isothermal\footnote{We also explored in parallel different values for $q_1$ to assess its sensitivity in the fitting process (see the Appendix \ref{sec:sensitivity})} ($q_1 =0$ thus T=constant), following the results of \citet{2018Kluska}. We set the inner radius to $R_\mathrm{in,~1}$ = 0.35~au \citep{Monnier2002}. Moreover, with the uncertainty on the nature of the hot material (T$\sim$1800 K) producing such an extended NIR emission, as raised by \citet{2018Kluska}, the emissivity factor $\epsilon_{\tau} = 1 - e^{(-\tau_{r, \lambda}/\cos{i})}$ was assumed to be achromatic and radially constant $(p_1=0$ and $\Sigma_{r}$= constant) for the inner disk component. 

For the outer component, we considered a temperature-gradient (and surface density-gradient) disc. We associated a standard dust composition to that outer component: a mixture of 70$\%$ amorphous silicates and 30$\%$ amorphous carbon, with the dust grain size distribution ranging between 0.05--3000 $\mathrm{\mu m}$, adopting a power-law of the MRN dust grain size distribution: $n \propto a^{-3.5}$. The opacities were computed with the DIANA tool \footnote{ Available at \href{http://dianaproject.wp.st-andrews.ac.uk}{http://dianaproject.wp.st-andrews.ac.uk}}. We fixed the exponent of the dust surface density profile for the outer component to $p_{2}$ = 1.5, as derived from the minimum mass solar nebula \citep{1977Weidenschilling}. 

As for the SED, we decided to only utilize the NIR and MIR parts, i.e. from 1--13~$\mathrm{\mu m}$, since our interferometric measurements end with the N-band. 

The list of free parameters of the two-component model are described in Table~\ref{tab:temp-grad results} as well as the priors ranges used for initiating the MCMC process. For several parameters, we could narrow down the priors range before running our full MCMC search in the following way:
\begin{itemize}
 \item we limited the range of explored values for the outer edge $R_\mathrm{{out, ~1}}$ of the inner component to 7--12~au based on the geometrical modeling results.
 \item concerning the inner and outer radii of the outer component: we made an initial guess on the different outer component parameters that reproduced roughly the MIR SED level. From that, we manually scanned a range of values (starting from 7~au) for the inner edge of the second component $R_{in, ~ 2}$ to match roughly the shortest baseline visibility level ($\sim$ 0.8). The same was done with $R_\mathrm{out,~ 2}$, considering the longer baselines that contribute to the second visibility lobe. Given the corresponding visibility level of 0.4, the outer edge of the outer component should not be too far from the inner edge, otherwise it would be too resolved, i.e. lower than 0.4. 
\end{itemize}


 \begin{figure*}
 \centering
 \includegraphics[width=17cm]{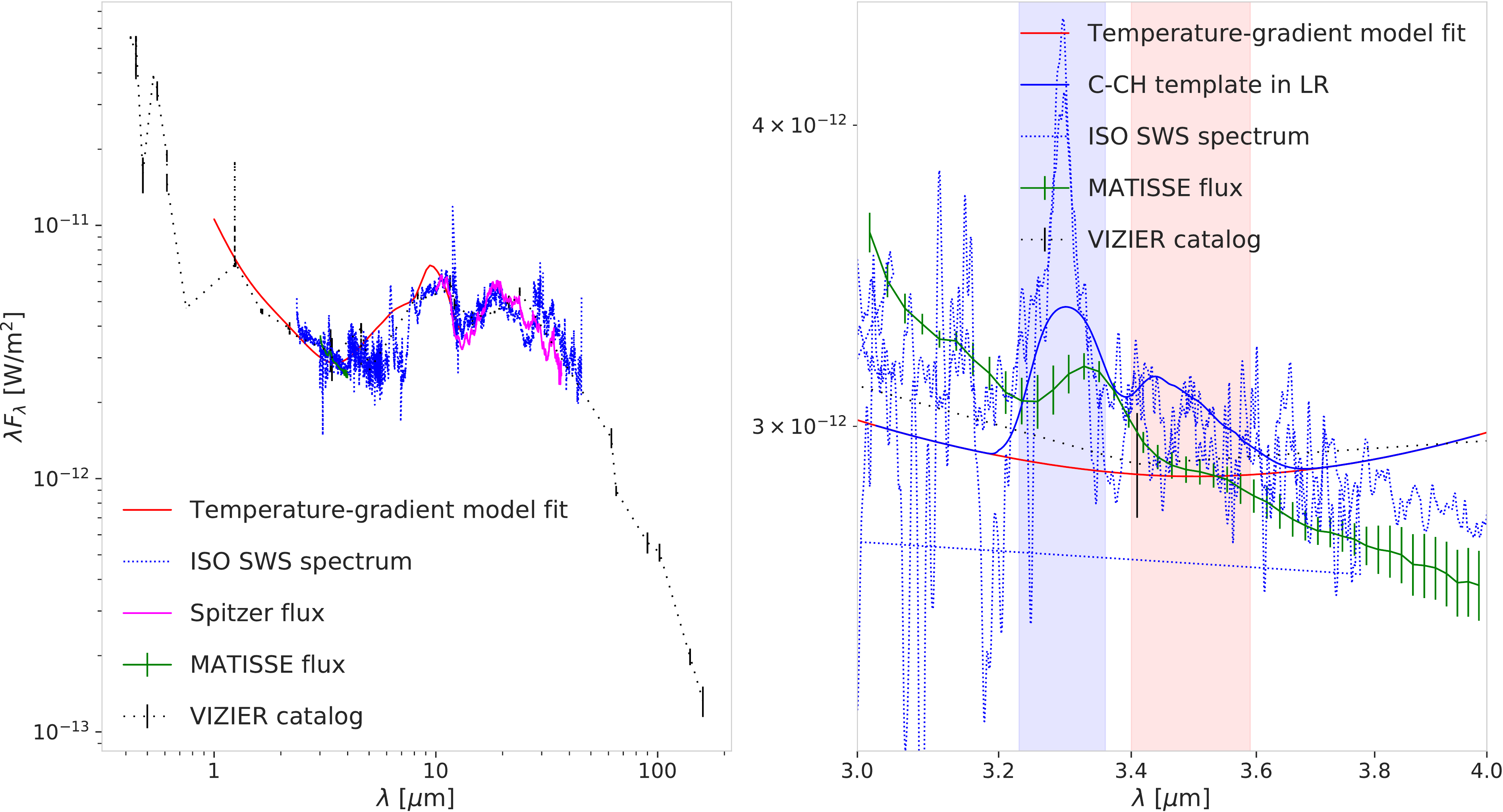}
 \caption{Left panel: SED from VIZIER database (dotted black line) with the fit for the temperature-gradient model (red line), including Spitzer (solid magenta line) and ISO SWS data (blue line) and MATISSE calibrated flux in green. On the right panel - zoom in to 3--4~$\mathrm{\mu m}$, where we represent the C-CH template in low resolution (blue line). Blue and red areas correspond to two different bands: 3.3~$\mathrm{\mu m}$ - aromatic and 3.4~$\mathrm{\mu m}$ - aliphatic. }
 \label{fig:SEDs}
\end{figure*}

\subsubsection{Modeling results}\label{subsubsection:results}

\begin{figure*}
 \centering
 \includegraphics[width=17cm]{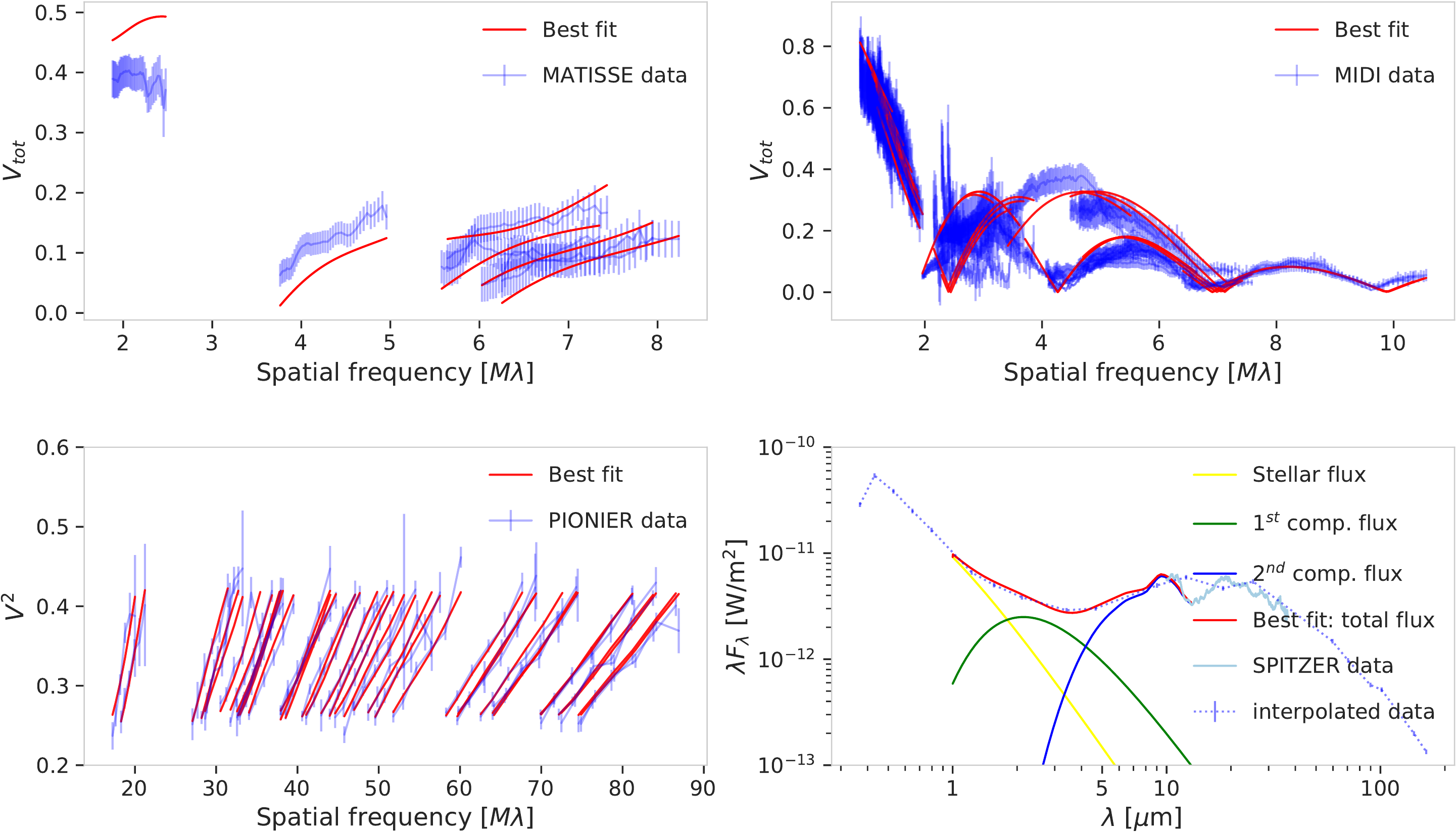}   
  \caption{Best temperature-gradient model fit of the extensive data set: PIONIER, MATISSE, MIDI and SED. The first panel row shows total visibilities vs spatial frequencies, while the second row shows first squared visibilities of PIONIER against spatial frequencies followed by the next panel with the SED. In the SED panel, there are different components in different colors which contribute to the total flux (red line). The red line in all panels represents the best-fit, while the blue one represents the data. In the SED panel three components are shown in different colors: yellow for the star, green - inner disk component and dark blue - outer component.}
 \label{Fig: all temperature gradient fits}
 \end{figure*}

\begin{figure*}
 \centering
 \includegraphics[width=\textwidth]{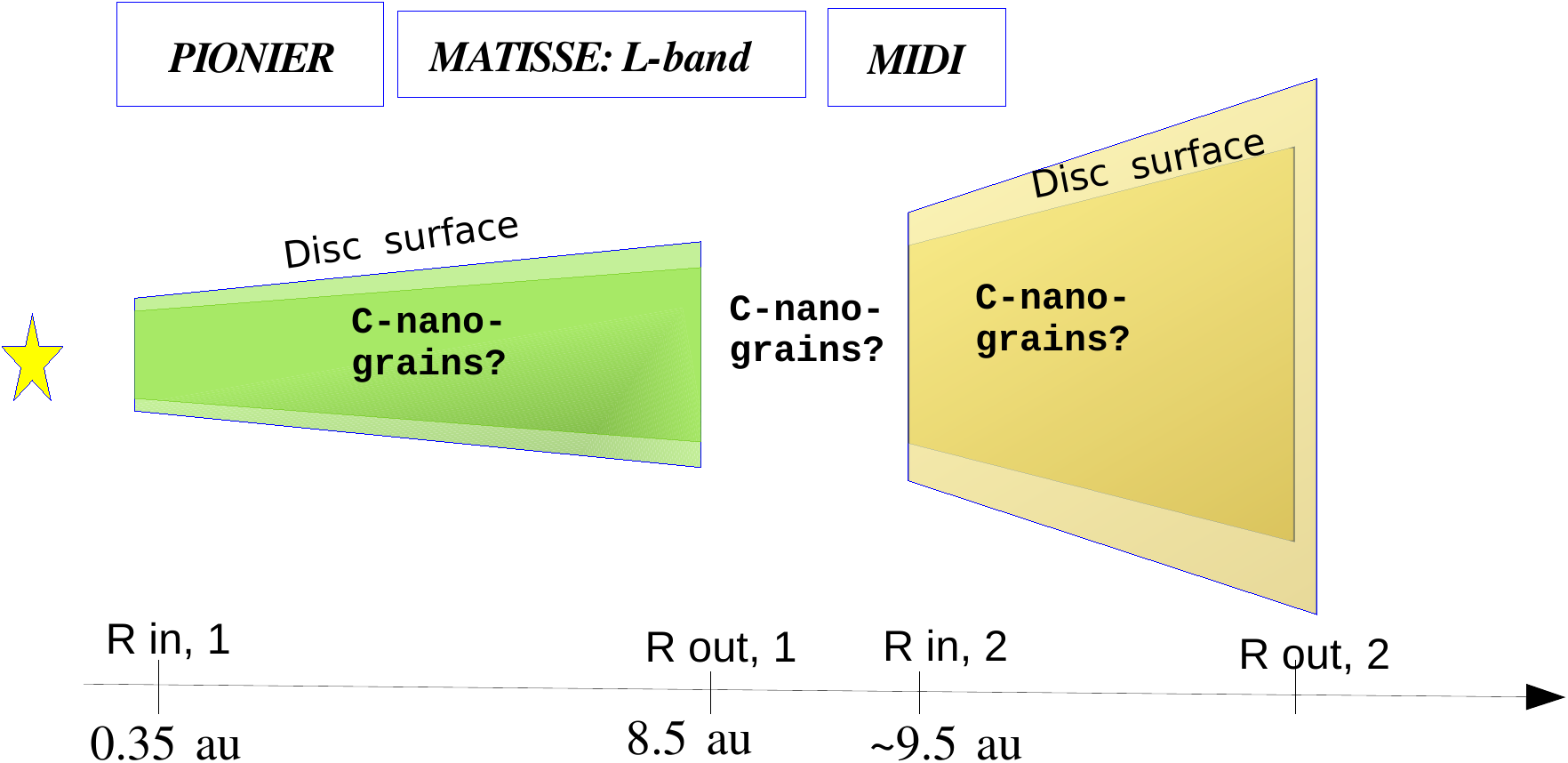}
  \caption{Schematics of HD\,179218. It shows a star and two disk components in different colors.}
 \label{Fig: Scetch of HD179218}
 \end{figure*}

Figure~\ref{Fig: all temperature gradient fits} shows that our simple approach fits nicely the various kinds of observations.
Together with the set of fixed parameters, the best-fit values for the free parameters are shown in Table~\ref{tab:temp-grad results}. The two-dimensional projections of posterior distributions of each free parameter are shown in Fig.~\ref{fig:temperature-gradient}.
Our best-fit model (See Fig.~\ref{Fig: Scetch of HD179218} for illustration) thus includes a hot isothermal inner component, which is extending up to $\sim$ 8.5~au and has a temperature around 1700~K. That temperature value is consistent with the previous results from \citet{2018Kluska} based on NIR interferometric data.
We discuss in Section~\ref{sec:discussion} whether stochastic heating of very small grains is a necessary condition to obtain such high temperatures far out in the disk at a distance of several au.
The median value of the emissivity factor, which is assumed achromatic and radially constant here (see Section~\ref{subsec:fitting}), approaches $4.7 \cdot 10^{-4}$. According to Eq.~\ref{eq:emissivity factor}, such a value translates to a vertical optical depth (thus also achromatic and radially constant) on the order of $10^{-4}$. The inner region is thus optically thin in the IR along our line of sight considering that the disk is not too much inclined. 
The outer disk component, starting at 9.5~au, is slightly larger than the outer radius of the inner component ($\sim$8.5~au). However, given the associated error bars, no conclusion can be drawn about the presence of a gap between the inner and outer components. Yet our best-fit model highlights a clear temperature difference. Indeed, the best-fit inner temperature of the outer component is 482~K which is significantly lower than the 1700~K derived for the inner component. The equilibrium temperature of dust grains located at a distance $r$ from the star, and directly irradiated by it, is given by:
\begin{equation}
T_\mathrm{dust} = T_{\star} \frac{1}{\epsilon^{1/4}} \sqrt{R_{\star}/2r} 
\end{equation}
where $\epsilon$ is defined by:
\begin{equation}
 \epsilon 	\equiv \frac{\int_{0}^{\infty} \kappa_{\lambda} B_{\lambda}(T_{dust})d \lambda /T^4_{dust}}{\int_{0}^{\infty} \kappa_{\lambda} B_{\lambda}(T_{\star})d \lambda /T^4_{\star}}
\end{equation}
In the case of small dust particles having a cooling efficiency $\epsilon < 1$, the heating is more efficient than cooling and the grains can be heated to temperatures higher than those of larger 'gray' grains ($\epsilon = 1$). By assuming $\epsilon = 0.1$, which is an average value for sub-$\mu$m-sized and $\mu$m-sized silicate grains, the upper limit on the expected dust temperature at a distance 9.5~au is 505~K, which is consistent with our best-fit inner temperature. 
In parallel, the mass of the outer dust component was found to be $4 \cdot 10^{-8}$~M$_\mathrm{\odot}$. In the frame of our model fitting where only IR data are considered, this mass estimate should mostly reflect the content of warm small sub-$\mathrm{\mu m}$ and $\mathrm{\mu m}$-sized grains in the atmosphere of the outer disk component; small sub-$\mathrm{\mu m}$ and $\mathrm{\mu m}$-sized grains usually dominating most of the disk opacity in the IR. Given the assumed surface density power-law and dust mass absorption coefficient for the outer disc, such mass translates to a vertical optical depth of 0.14 at 10~$\mathrm{\mu m}$ at the outer disk inner radius. 

 According to the corner plots displayed in Fig.~\ref{fig:temperature-gradient}, most parameters are fairly well constrained (i.e. associated with a peaked posterior probability distribution) with the exception of the outer edge and the temperature power-law exponent of the outer component. The sensitivity to the parameters driving the 'large-scale' structure of the outer disk component is thus weak. It is expected since our modeling was based on IR interferometric data and the short wavelength side (< 15~micron) of the SED, which are more sensitive to the parameters driving the warm and smaller scales in the inner regions, on which we focus our attention. 
 As a final note on the modeling results, our best-fit model slightly overestimates the lowest frequency MATISSE visibility while underestimating the following one represented by medium spatial frequencies (see Fig.~\ref{Fig: all temperature gradient fits}). As previously shown by our geometrical modeling (see Section \ref{subsec:geom_mod_results}), the L-band and the NIR bands are mainly sensitive to the inner disc.
 However, contrary to the PIONIER visibilities, which cover only high spatial frequencies and thus overresolved the inner disk emission, the MATISSE ones partially cover the low-frequency part of HD\,179218's visibility function. That makes them more sensitive to the global size and geometry of the inner component. Such a discrepancy with the low-frequency MATISSE visibilities thus indicates that our simple isothermal model cannot fully reproduce the characteristics of the inner component emission. In the Appendix \ref{sec:sensitivity}, we explore the sensitivity of several inner disk parameters, to identify possibly the reason for such a discrepancy and improve our agreement with the low-frequency MATISSE visibilities.

\begin{table*}\centering
\captionof{table}{Fitting parameters and priors for temperature-gradient model. Median fit parameters for the one component models together with $1\sigma$ error bars; reduced $\chi^2$ corresponds to the best-fit; inclination and position angle are fixed at $i$ = 54$^{\circ}$, $\theta$ = 26$^{\circ}$.}
\begin{tabular}{c c c c c}
\hline\hline
 \multicolumn{5}{c}{Isothermal + Temperature-gradient Model} \\
\multicolumn{5}{c}{PIONIER + MATISSE + MIDI + SED }\\
\hline
 reduced $\chi^2$ & & & &4.6 \\
\hline
Description & Parameter & Prior & Median & Best-fit \\
Source distance & $d~ [\mathrm{pc}]$ & 266 & / & /\\
Effective temperature & $T_\mathrm{eff} ~[\mathrm{K}]$ & 9500&/&/ \\
Stellar radius & $R_{\star} ~[R_\mathrm{\odot}]$ & 3.5 & / &/ \\
\hline

Inner edge of the $1^{st}$ component &$R\mathrm{_{in, ~1 }} ~[\mathrm{au}]$& 0.35 &/&/ \\
Temperature power-law of the $1^{st}$ component & $q_{1}$ & 0.0 &/&/ \\
Surface density power-law of the $1^{st}$ component & $p_1 $& 0.0 &/&/ \\
Surface density power-law of the $2^{nd}$ component& $p_2$ & 1.5& /&/ \\
 \hline
 Emissivity factor & $\epsilon_{\tau_{1}} \cdot 10^{-4}$ & from 0.1 to 10 & 4.7$^{+1.7}_{-1.6}$ &4.93 \\
Temperature of the isothermal inner component& $T_\mathrm{in, ~1}~[\mathrm{K}]$ & from 1500 to 1800 & $1712^{+114}_{-98}$ &1714 \\
Outer edge of the $1^{st}$ component&$R_\mathrm{out, ~1}~[\mathrm{au}]$ &from 7 to 12 & $8.67^{+0.96}_{-0.76}$ & 8.5 \\
Mass of the $2^{nd}$ component& $M_\mathrm{2} \cdot 10^{-8}[\mathrm{M_{\odot}}]$ & from 0.5 to 10 &3.3$^{+1.5}_{-1.3}$ &4.0\\
Temperature at the inner edge of $2^{nd}$ component &$T_\mathrm{in, ~2}~[\mathrm{K}]$ & from 200 to 800 & $479.7^{+47}_{-41}$ & 482 \\
Inner edge of the $2^{nd}$ component&$R_\mathrm{in, ~2}~[\mathrm{au}]$ & from 8 to 12 & $9.48^{+0.84}_{-0.78}$ & 9.45 \\
Outer edge of the $2^{nd}$ component&$R_\mathrm{out, ~2}~[\mathrm{au}]$ & from 12 to 16 & $14.97^{+0.67}_{-0.6}$ & 15.3\\
Temperature power-law of the $2^{nd}$ component &$q_{2}$ & from 0.5 to 1& $0.73^{+0.16}_{-0.15}$& 0.92\\
 \hline
\end{tabular}
\label{tab:temp-grad results}
\end{table*}

\section{Carbonaceous features in the L-band data} \label{sec: PAHs}
In the previous section, we built a disk model (see Fig.~\ref{Fig: Scetch of HD179218}) that reproduces in a satisfactory way the IR continuum emission of HD\,179218 as represented both in the spectrum and in the visibilities from several instruments. We now focus on a deeper characterization of the solid material, especially its potential carbonaceous nature. Indeed, carbonaceous features (aromatic and aliphatic) were detected between 0\farcs1 -- 0\farcs5 
(30 - 120 au) from the star with the VLT/NACO instrument \citep{2019Bouteraon}. Due to a lack of angular resolution, no information could be provided on the regions $< 30$~au. One of the ambitions of our work is to check if the extended inner emission, attributed to stochastically heated particles, can come from solid carbonaceous nanometric species like PAHs. 
We aim: 1) to determine if features associated with carbonaceous nano-grains are already detectable in our low resolution ($R \sim$30) L-band MATISSE data (particularly the aromatic feature at 3.3~$\mathrm{\mu m}$); and 2) to assess the feasibility of detection and characterization (position, size, rough composition) of those carbonaceous nano-particles in the inner 10~au region from future MATISSE observations in medium spectral resolution ($R \sim$500).
\subsection{Adding a carbonaceous component to the model} 
To investigate the presence of carbonaceous features in our MATISSE data, we added an additional ring component to our continuum disk model described in Section \ref{sec:continuum} to represent this emission. We associated with that ring an emission template containing various aromatic and aliphatic features derived 
from a decomposition of L-band spectra taken with the VLT/NACO instrument (see our Fig.~2 and Fig.~3 in \citet{2019Bouteraon}). 

Following Section \ref{sec:continuum}, we note the total continuum flux of our best-fit temperature-gradient model at a given wavelength~$\lambda$ is $F_{\rm cont}(\lambda)=F_{\lambda,~ star}+F_{\lambda,~ inner ~disc}+F_{\lambda,~ outer~ disc}$. Then we include an additional component, named hereafter C-CH disc, that emits only in the wavelength range defined by the template from $\sim 3.2 -3.6~\mathrm{\mu m} $. That C-CH disk extends from $r_{\rm in, ~C-CH}$ to $r_{\rm out, ~C-CH}$, and has a flux ratio of $f_{C-CH/Cont}(\lambda)$ with respect to the continuum emission. The total flux ratio is expressed as: $f_{C-CH/Cont}(\lambda)=f_{C-CH/Cont}(3.3 \mu {\rm m}) \cdot t(\lambda)$, where $t(\lambda)$ is the normalized carbonaceous template extracted from the NACO's observations; the 3.3~$\mathrm{\mu m}$ wavelength corresponds to the position of the strongest feature in the emission template, i.e. the aromatic-CH one. The parameters of our C-CH disk component are thus $f_{C-CH/Cont}(3.3 \mu {\rm m})$, $r_{\rm in, C-CH}$ and $r_{\rm out, C-CH}$. 

At a given wavelength in the range explored by MATISSE ($3 - 4 ~\mathrm{\mu m}$), we calculate the flux for the C-CH disc:
\begin{equation}
F_{\rm C-CH}(\lambda)=\frac{\cos i}{d^2}\int_{r_{\rm in, C-CH}}^{r_{\rm out, C-CH}}I_{\lambda}(r)\:2\pi\:r\:dr,
\end{equation}
where $I_{\lambda}(r)$ is the intensity with which each infinitesimal ring composing the C-CH disk will emit. In this study we consider an intensity that follows a $r^{-2}$ intensity radial profile. 
This is motivated by the fact that the carriers are being stochastically heated and the band intensity is expected to be proportional to the strength of the far-UV radiation field. The 1/$r^{2}$ law corresponds to the dilution law of the far-UV radiation field strength. 
The above described band intensity behavior has already been observed in several protoplanetary discs in the L-and N-bands at intermediate to large distances (40-300~au) \citep[e.g.,][]{2006Lagage, 2020Habart} (see Figs.~3 and 12 of the latter) and explained the spatial extension of the bands.
In the case of HD\,179218, we computed from the NACO/VLT data presented in \citet{2019Bouteraon} the spatial emission profile of the aromatic band at 3.3~$\mathrm{\mu m}$. As shown in Fig.~\ref{fig:intensity}, it is at a first order proportional to $r^{-2}$ at distances of 40 to 120~au.
Without further constraints on the intensity profile the aromatic and aliphatic bands inside 40~au, we thus assumed the same radial dependency across the whole disc: $I_{\lambda}(r)=I_{\lambda,in}(r/r_{\rm in,C})^{-2}$. Given that a constant intensity profile is favored for the IR continuum emission in the inner 10~au region (see Section~\ref{sec:continuum}), it means that we treat separately, in a first approximation, the grain populations responsible for the continuum and the aromatic/aliphatic bands. We describe in Section~\ref{sec:new intensity prof} the effect of a constant intensity profile for the aromatic/aliphatic bands in the inner 10~au region, thus following the IR continuum emission.  
The total flux of the C-CH disk is :
\begin{equation}
F_{\rm C-CH}(\lambda) = f_{\rm C-CH/Cont}(3.3 \mathrm{\mu m}) \cdot t(\lambda) \cdot F_{\rm cont}(\lambda)
\end{equation}
where $t(\lambda)$ is the template introduced above. This leads to :
\begin{multline}
\frac{\cos i}{d^2}\int_{r_{\rm in,C-CH}}^{r_{\rm out,C-CH}}I_{\lambda,in} \left(\frac{r}{r_{in,C-CH}}\right)^{-2}2\pi\:r\: dr=\\ =f_{\rm C-CH/Cont}(3.3 \mathrm{\mu m}) \cdot t(\lambda) \cdot F_{\rm cont}(\lambda)
\end{multline}

Thus $I_{\lambda,in}$ is:
\begin{equation}
 I_{\lambda,in} = \frac{f_{\rm C-CH/Cont}(3.3 \mathrm{\mu m}) \cdot t(\lambda) \cdot F_{\rm cont}(\lambda)}{\frac{2 \pi ~\cos i \:r_{in,C-CH} \ln{\left( \frac{r_{out,C-CH}}{r_{in,C-CH}}\right) } }{d^2}} 
\end{equation}
The correlated flux produced by the C-CH disk with scaled intensity will become:
\begin{multline}\label{eq:PAH corr flux uniform}
f_{\rm corr,\lambda, C-CH}(B_{\rm p}(i,\theta))\\
=\frac{2\pi \cos i}{d^2}\int_{r_{\rm in}}^{r_{\rm out}}I_{\lambda,in}\:\:J_0\left[\frac{2\pi}{\lambda}B_{\rm p}(i,\theta)\frac{r}{d}\right]\:r \left(\frac{r}{r_{in,C-CH}}\right)^{-2} dr 
\end{multline}

\subsection{modeling approach} \label{subsec:30per_mod_approach}
With our new model combining the best-fit continuum model with a C-CH disk component containing the carbonaceous features, we produced several L-band observables:
\begin{itemize}
 \item a low spectral resolution total spectrum that we directly compared to the calibrated total spectrum of MATISSE to identify possibly the aromatic feature at 3.3~$\mu$m. 
 \item low spectral resolution ($R$=30) visibilities that we directly compared to the measured MATISSE visibilities to determine if first constraints can already be obtained on the region emitting the carbonaceous features (especially the aromatic feature at 3.3~$\mu$m), especially its inner radius. 
 \item medium spectral resolution ($R$=500) visibilities, associated with error bars, to simulate future MATISSE observations in medium spectral resolution. The error bars were generated using an IDL MATISSE simulation tool, which is based on a noise model detailed in \citet{2016SPIE.9907E..28M}. The noise level was computed following the L-band flux of HD\,179218.
\end{itemize}
 
We set the flux contribution from the C-CH disk $f_{C-CH/Cont}(3.3 \mu {\rm m})$ to 30\% of the continuum, as obtained with the VLT/ISAAC instrument \citep{2006Acke}. Such a choice is justified by the fact that: 1) VLT/ISAAC and MATISSE have a FOV of the same order: 1 arcsec for VLT/ISAAC and 0.6 arcsec for MATISSE (in L-band with ATs); 2) there are no other measurements, with smaller FOVs, on the 3.3~$\mu {\rm m}$ feature-to-continuum ratio of HD\,179218. A 30\% band-to-continuum ratio thus seems a reasonable starting point for the modeling. 
The outer radius was fixed to 80~au, based on the FOV of MATISSE in L-band with the ATs. This FOV is set by the angular diameter of the MATISSE spatial filter device (pinhole), which is 1.5$\lambda/D$ at $\lambda=3.5$~$\mu$m ($\simeq 600$~mas, i.e., 156~au at the distance of 266~pc).

From that, we thus explored different values of the inner radius $r_{in,C-CH}$, starting from 0.35 up to 30~au. The results are described in the following subsection.

\subsection{Modeling results in low and medium spectral resolution}



\begin{figure*}
 \centering
  \includegraphics[width=17cm]{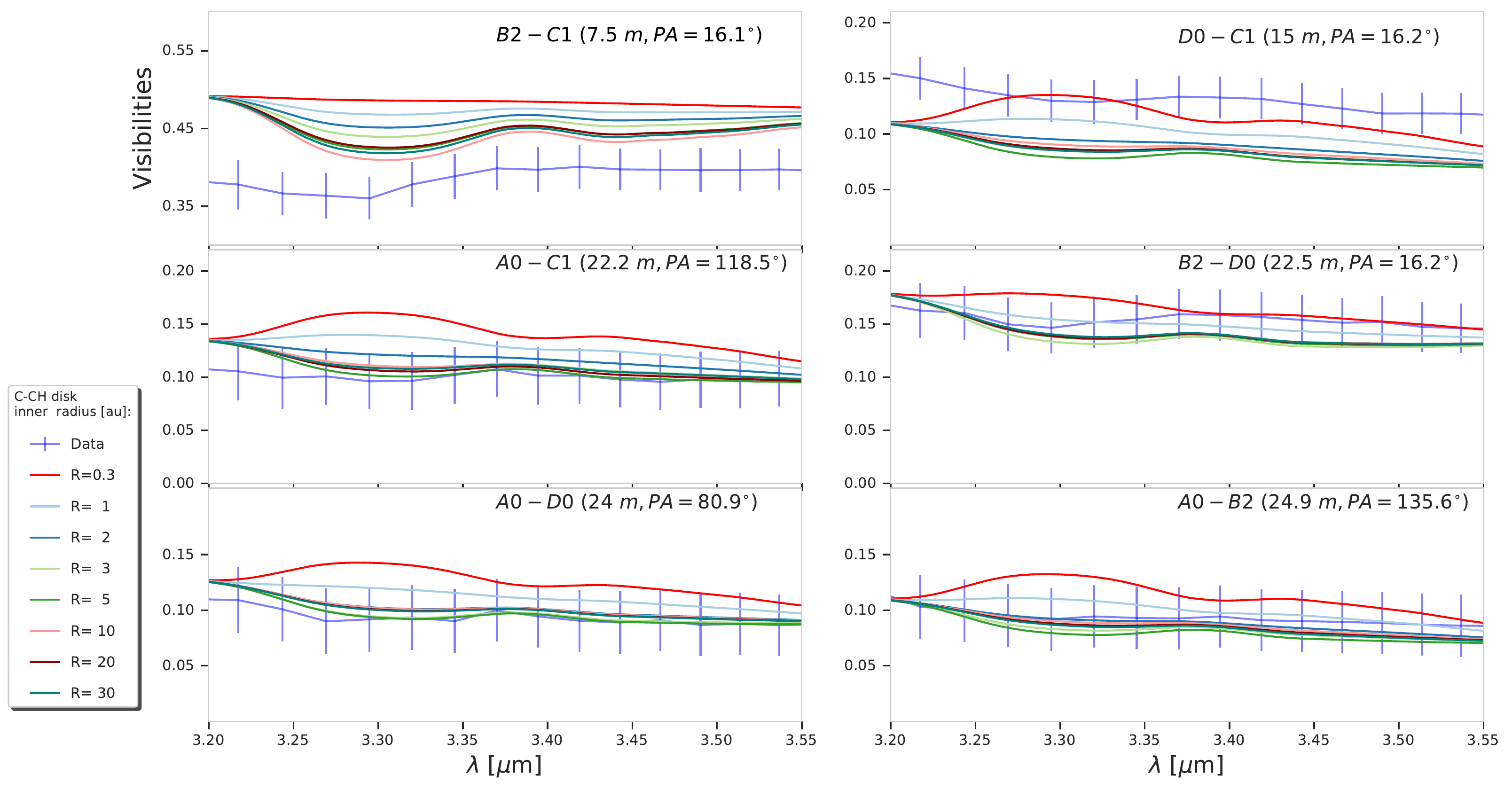}
 \caption{Low spectral resolution ($R$=30) L-band visibilities of the model combining the best-fit continuum model, from Section~\ref{subsubsection:results}, with a C-CH disk component. Different inner radii of the C-CH disk are explored - see the legend on the left hand side. The telescope pairs, projected baseline length ($B$) and the position angles of the projected baseline ($B_{PA}$) are shown in each subplot. The measured L-band MATISSE visibilities are overplotted in light blue solid line (with error bars).}
 \label{fig:scaled PAH in LR}
\end{figure*}

\begin{figure*}
 \centering
 \includegraphics[width=17cm]{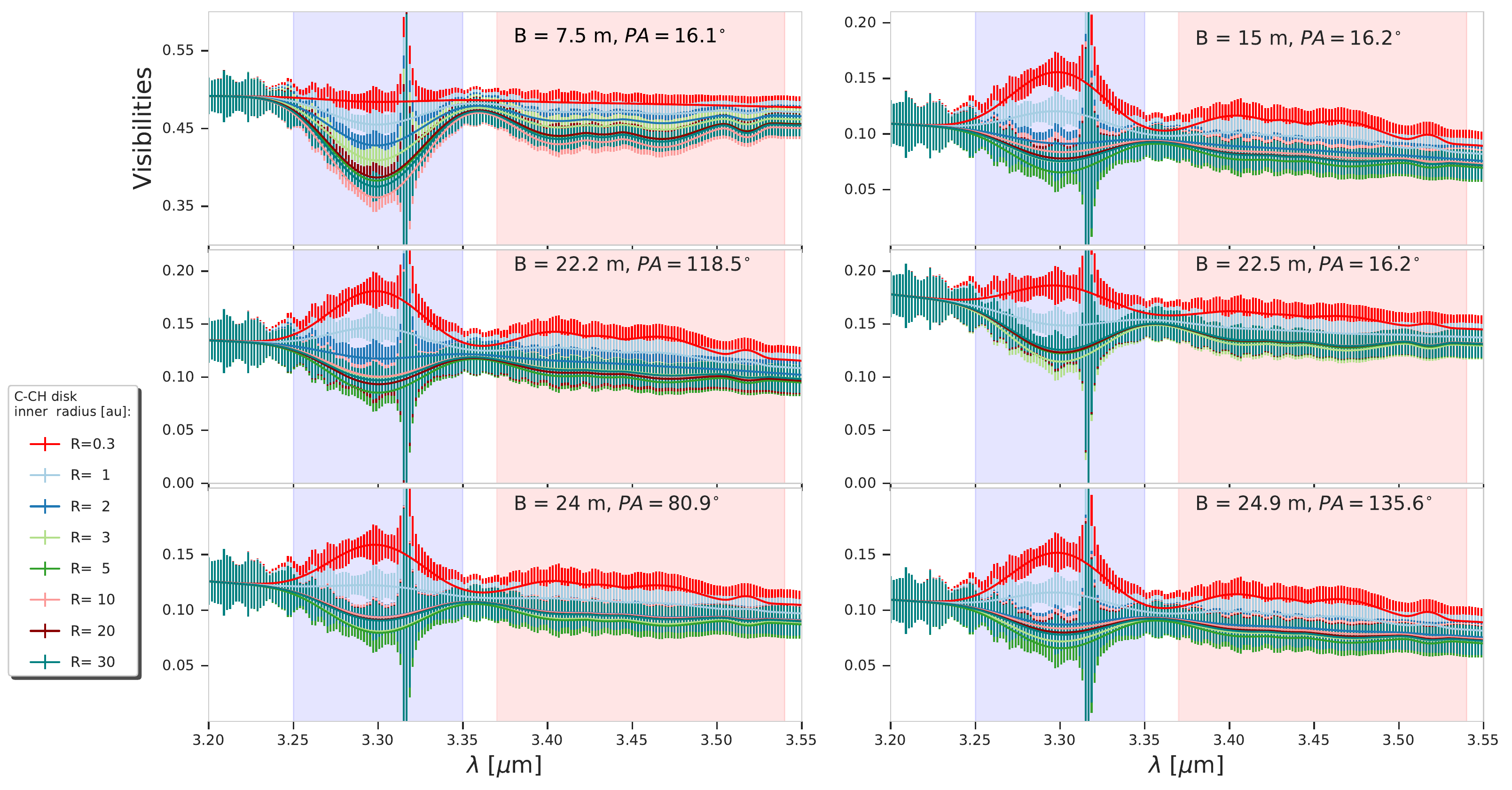}
 \caption{Simulated medium resolution ($R$=500) MATISSE visibilities based on the model combining the best-fit continuum model from Section~\ref{subsubsection:results}, with a C-CH disk component. Different inner radii of the C-CH disk are explored - see the legend on the left hand side. The telescope pairs, projected baseline length ($B$) and the position angles of the projected baseline ($B_{PA}$) are shown in each subplot. Blue and red areas correspond to two different bands: 3.3~$\mathrm{\mu m}$ - aromatic and 3.4~$\mathrm{\mu m}$ - aliphatic. The expected error bars were computed from a noise model of MATISSE that includes a realistic atmospheric transmission profile (see details in the text); note that the larger error bars around 3.31~$\mathrm{\mu m}$ comes from a strong telluric water vapor absorption line.}
 \label{fig:scaled PAH in MR}
\end{figure*}

As shown in Fig.~\ref{fig:SEDs}, the model L-band total spectrum presents a noticeable bump due to the aromatic feature around 3.3~$\mathrm{\mu m}$, even though it is attenuated by the lower spectral resolution.
This bump around 3.3~$\mathrm{\mu m}$ coincides with a strong telluric line at the same wavelength. 
That makes our comparison not conclusive about the detection of the 3.3~$\mathrm{\mu m}$ feature in the total spectrum of MATISSE. Nevertheless, knowing that aromatic and aliphatic features were detected down to 30~au by NACO \citep{2019Bouteraon} and that the MATISSE FOV is a radius of about 80~au (from the central star), it is unlikely that the aromatic features completely faded out in the MATISSE spectrum since the aromatic rings are more resistant to UV processing than aliphatic bonds \citep{2019Bouteraon}. So we expect them to be present in the future higher spectral resolutions MATISSE spectra.

Figure~\ref{fig:scaled PAH in LR} shows the low resolution model visibilities for different assumed inner radius values of the C-CH disc, overplotted on the measured visibilities. The comparison of the low spectral resolution visibility data with the model visibilities, which assumes a contribution of 30$\%$ from the C-CH disk at 3.3~$\mathrm{\mu m}$, suggests that the region emitting the carbonaceous features cannot start from within a too small inner radii ($\leq$ 1 au). Indeed, such models produce a visibility excess around 3.3 $\mathrm{\mu m}$, for most of the baselines, that is not seen in the measured visibilities. 
The emitting regions starting from larger radii, but still within the inner 10~au (i.e. inside the isothermal inner component of our best-fit continuum model), produce a slight visibility drop around 3.3~$\mathrm{\mu m}$, which remains consistent with the measured visibilities. Therefore, the 2-10~au range of inner radius values could contain nano-grain components of carbonaceous nature. 
For even larger inner radii ($>$ 10~au), the model visibilities are being almost indistinguishable from each other on all baselines except marginally for the smallest one.

Figure~\ref{fig:scaled PAH in MR} shows the model visibilities in medium spectral resolution with the expected error bars of MATISSE in the case of HD\,179218. Again, different inner radius values for the C-CH disk were explored. Different inner radii could be discriminated at the scale of an au in medium spectral resolution data. Indeed, the simulated visibilities, corresponding to inner radii of 1~au, 2~au, and 3~au, show significant differences around the 3.3~$\mathrm{\mu m}$ feature, i.e. larger than the predicted visibility error bars. For larger inner radii, such as the considered values of 5, 10, 20 and 30~au, the simulated visibilities overlap within the error bars and cannot be distinguished. With medium resolution observations, an au-scale precision determination of the inner radius of the 3.3~$\mathrm{\mu m}$ aromatic feature emitting region thus seems possible if it lies in the first few au. If the radius turns out to be larger, only a lower limit could then be provided. 
Going to medium resolution will also help to better analyze the relative variations of different carbonaceous features: aromatic - 3.3 $\mathrm{\mu m}$ versus aliphatic - 3.4--3.6 $\mathrm{\mu m}$. These two groups of features can be clearly separated according to the display of Fig.~\ref{fig:scaled PAH in MR}.\\

Of course, those conclusions are based on our initial assumption of a global 3.3$\mu {\rm m}$ band-to-continuum ratio of 30\% ($f_{C-CH/Cont}(3.3 \mu {\rm m})$=30\%). The possibility of a radially varying band-to-continuum ratio cannot be excluded, as shown for instance in N-band in the case of HD100546 \citep{2020Habart}. Nevertheless, the influence of a locally lower or higher band-to-continuum ratio is rather straightforward: a smaller ratio implies a harder detection of the 3.3~$\mu {\rm m}$ feature and the other way around. We made several tests with lower band-to-continuum ratios. In particular, with $f_{C-CH/Cont}(3.3 \mu {\rm m}) < 10$\%, the different model visibilities (associated with the different inner radii of the C-CH disk), become all consistent with the LR MATISSE visibilities within the error bars.

\subsection{A flat intensity profile in the inner 10 au region} \label{sec:new intensity prof}
In that section, we considered a global intensity profile $r^{-2}$ for the C-CH disc. However, that may change in the inner regions, especially since a flat intensity profile is favored for the IR continuum emission (see Section~\ref{sec:continuum}). Moreover, the aromatic band intensity profile of HD\,179218 presented in Fig.\ref{fig:intensity}) seems to show a slight deviation from the $r^{-2}$ power-law fit inside $\sim$ 55~au. Assuming that the aromatic/aliphatic band carriers are also producing the IR continuum emission, we tested the case of a flat intensity profile for the C-CH disk up to 10~au and a $r^{-2}$ intensity profile from $\sim$ 11~au. As shown in Appendix~\ref{fig:LR_new_intensity}, it would not be a favorable case for MATISSE. Indeed the different model visibilities (associated with the different inner radii of the C-CH disc) are all consistent with the LR MATISSE visibilities. Moreover, they show a deficit around 3.3~$\mu$m for all baselines, which indicates that the C-CH disk is overresolved even for the smallest baseline length. No clear constraint can be provided on the inner radius of the aromatic/aliphatic bands emitting region. Nevertheless, it appears that having a global intensity profile in $r^{-2}$ or a locally constant one (in the inner region) imprint clear differences in the behavior of the model MATISSE visibilities. Future MATISSE MR data may thus help to distinguish between the two different intensity profiles.

\section{Discussion}\label{sec:discussion}
\subsection{Stochastically heated grains in the inner 10~au region?} \label{subsection: SHGs}

Our temperature-gradient modeling favors a two-component disk structure. 
These two components trace two distinct regions at about 10~au. Notably, a hot ($\sim$ 1700~K) and isothermal inner 10~au region (see Fig.~\ref{Fig: Scetch of HD179218}) is required to reproduce the SED, the existing near-infrared interferometric data, and the L-band MATISSE visibilities. 



\begin{figure*}
\sidecaption
  \includegraphics[width=12cm]{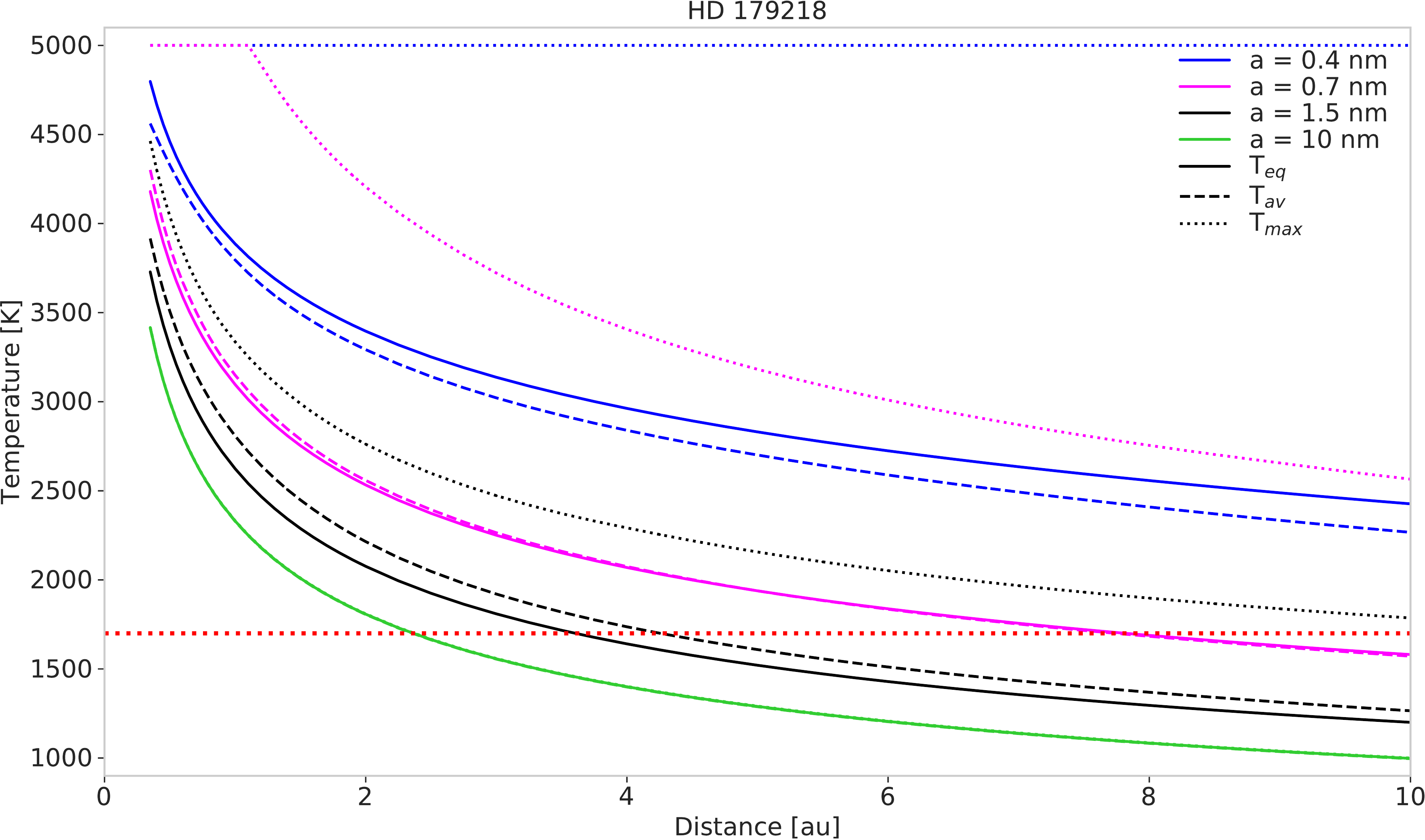}
     \caption{Different temperature indicators: $T_{eq}$, $T_{av}$ and $T_{max}$ of amorphous carbonaceous nano-grains (mostly aromatic), taken from the THEMIS grain model, versus distance from the star. Different grain sizes are considered: 0.4, 0.7, 1.5 and 10~nm. The temperature indicators correspond to the equilibrium temperature, average temperature and maximum temperature, respectively. The red dotted line corresponds to 'model required' temperature of 1700~K. No sublimation or grain destruction was considered here.}
     \label{fig:themis_Tdust}
\end{figure*}

As mentioned in Section \ref{subsubsection:results}, stochastically heated very small grains could be invoked to explain such high temperatures up to 10~au distances from the central star. To illustrate quantitatively that, let us consider the case of carbon nano-grains. Indeed, they can reach very high temperatures in highly irradiated regions like the inner 10~au disk region around HD\,179218.
 Figure~\ref{fig:themis_Tdust} presents the radial profile of the average and maximum temperatures ($T_\mathrm{av}$ and $T_\mathrm{max}$), which can be reached by stochastically heated carbonaceous nano-grains up to 10~au in an optically thin environment similar to the inner region of HD\,179218's disc. We considered the amorphous carbonaceous nano-grains, mostly aromatic (low band gap material $\sim$ 0.1 eV, low hydrogenation), from the THEMIS model. THEMIS is a model characterizing the grain population: size, composition and structure. The temperatures were computed with the DUSTEM\footnote{\href{Available at}{https://www.ias.u-psud.fr/DUSTEM/}} tool, which calculates the temperature and emission of various dust species in the optically thin limit within a default spectral range between 0.04 and $ 10^5 \mathrm{\mu m}$. Also displayed in Fig.~\ref{fig:themis_Tdust}, $T_\mathrm{eq}$ is the theoretical equilibrium temperature: the temperature at which the absorbed and emitted energies are equal and corresponds to the emission temperature.

 

Independently from their lifetime, we can see that all grains can reach temperatures greater than 1000 K up to 10 au. For the largest $10~\mathrm{nm}$ grains, $T_\mathrm{av}$ and $T_\mathrm{max}$ converge to $T_\mathrm{eq}$ which means that the radiative equilibrium is reached for this grain size. As expected from other studies \citep[e.g.,][]{1992Siebenmorgen}, the bigger the grain is, the more likely it will reach thermal equilibrium. For such large grains, a temperature on the order of 1700~K can be reached at 2~au from the star. As expected, smaller nm-sized grains are subject to a significant stochastic heating with temperature excursions well above the equilibrium temperature. The $1.5~\mathrm{nm}$ and $0.7~\mathrm{nm}$ grains reach an average temperature of 1700~K even at far distances: $\sim 5$ and $\sim 10$ au, respectively. Finally, the smallest grains can experience extremely high temperatures (> 5000~K) while having average temperature higher than 2500~K, which makes their survival in the inner region of HD\,179218 very unlikely. 


 
\begin{figure*}
 \centering
 \includegraphics[width=\textwidth]{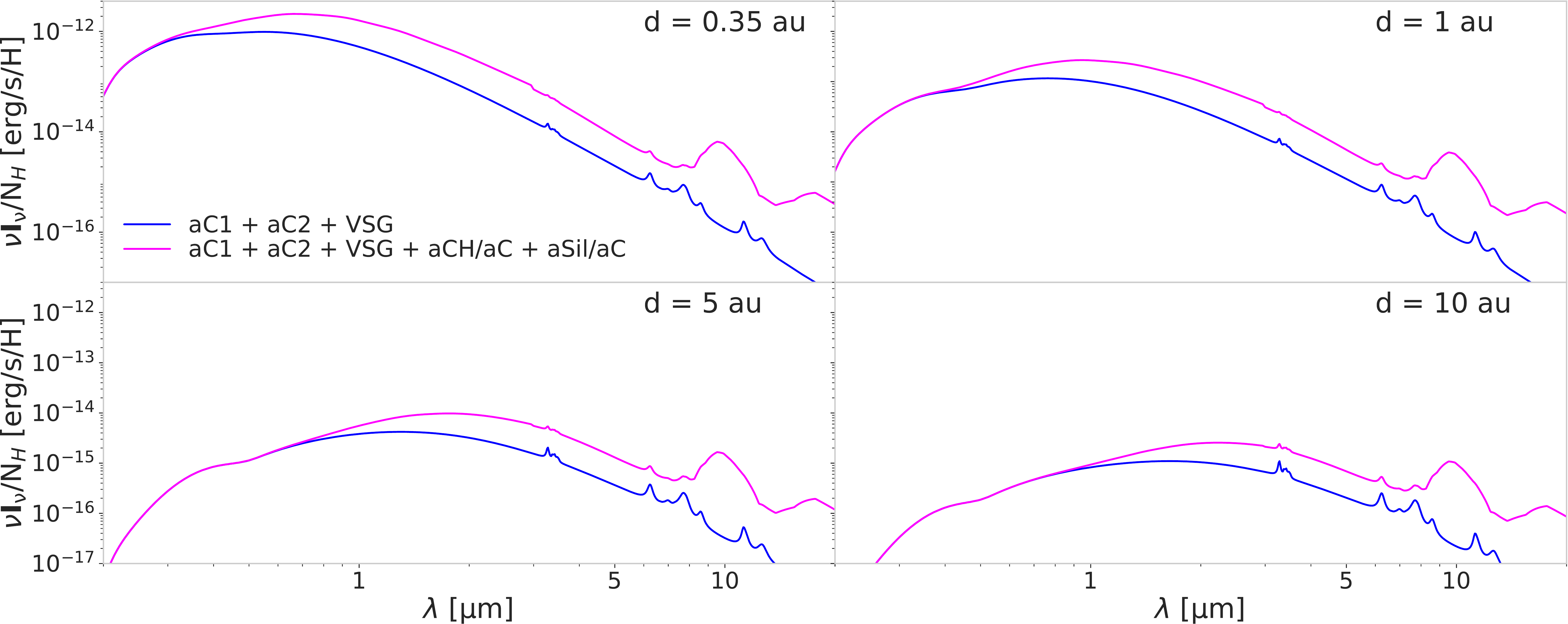}
 \caption{Thermal emission of the amorphous carbon nano-grains population (aC1, aC2, VSG) of the THEMIS model (blue), and of the entire THEMIS grain population including larger 100~nm-sized carbon (aCH/aC) and silicates (aSil/aC) grains (pink), for different distances to the central star. The SED units are erg/s/H which is the emission per proton for each grain population. See papers \citet{2020Habart} and \citet{2017Jones} for more detail on the different grain species considered in THEMIS. 
 }
 \label{fig:themis_SED}
\end{figure*}

Figure~\ref{fig:themis_SED} shows, at different distances from the star, the thermal emission of the carbon nano-grains population (blue line) of the THEMIS model, and the one of the entire grain population (pink line) of the THEMIS model, which also includes larger 100~nm-sized carbon and silicate grains \citep[see][]{2020Habart}. It appears that the smallest particles (blue line) produce a significant continuum emission in the NIR and in the L-band. Moreover, their contribution to the emission of the entire grain population is increasing at larger distances from the star. 
Then, by considering only the population of nano-grains, Fig. \ref{fig:themis_Tdust} indicates that a temperature of 1700~K is reached by 10~nm grains at 2~au, 1.5~nm grains at 4~au, and 0.7~nm grains at 8~au. A high temperature could thus be maintained over the inner 10~au region with the largest nano-grains setting the innermost condensation rim, and smaller hot grains being more and more present at larger distances. The smaller nano-grains would thus contribute to produce a high temperature level through their stochastic heating at large distances. 
In other words, the `isothermal' and hot nature of that inner 10~au region would be linked to the radial decrease of the minimum size of the nano-grains that can survive at a given distance from the star. 

 Aside from those conclusions based on thermal emission only, we may wonder if scattering (whether we refer to stellar light or dust emission self-scattering) could also contribute significantly to the inner 10~au emission. However, grains smaller than 100~nm have a very low scattering efficiency in the Rayleigh regime in the NIR and MIR \citep[see][]{2015Arai}. To illustrate the negligible scattering contribution from very small grains, we present in Appendix \ref{sec:sensitivity} a radiative transfer simulation using THEMIS grains. As a consequence, a significant contribution from scattering in the NIR and MIR would require the presence of larger sub-$\mathrm{\mu m}$-sized and $\mathrm{\mu m}$-sized grains. However, the presence of such large grains would induce a radial temperature gradient with temperatures well below what we find. The dominant brightness of the inner 10 au region, interpreted by the 'hot' nature of the small dust particles, is thus unlikely to be linked to scattering.

Another important result on the grain emission in the inner 10~au region of the HD\,179218's disk is that a flat (i.e., radially constant) intensity profile is favored by our data (see our best-fit temperature-gradient model in Section \ref{sec:continuum}). Several tests were performed in Appendix \ref{sec:sensitivity} to explore different surface density gradients. They confirmed that a flat intensity profile provides the best agreement with the data. Interestingly, a flat emission from nano-grains was also observed in HD\,100546 between 20 and 40~au \citep{2020Habart}.
Such a flat intensity profile, produced by an inner 10~au region that is optically thin both in our line of sight at IR wavelengths and radially at UV and visible wavelengths (see Appendix \ref{sec:sensitivity}), must be interpreted by considering the dilution proportionally to $r^{-2}$ of the stellar flux. Indeed, since the stellar radiation field decreases as $r^{-2}$, something should compensate for that flux dilution to maintain the intensity of the disk emission as we go further from the star. A positive gradient in the radial distribution of dust could be a possibility. Such radially increasing dust density could be linked to the inside-out destruction of the small dust particles together with local replenishment or from the outer regions. Destruction and replenishment are addressed in more details in the following subsection.

\subsection{Two separated grain populations}\label{sec:two grains}
Our temperature-gradient modeling, combined with the theoretical elements provided in Section \ref{subsection: SHGs}, points toward the presence of an optically thin (NIR vertical optical depth $\sim 10^{-4}$; see Section \ref{subsubsection:results}) population of hot nano-grains in the inner 10~au region. This is associated with an outer (r > 10~au) ring-like component of larger, probably $\mathrm{\mu m}$-sized, grains, which are responsible for the N-band emission.



Thus, a question arises: how can we have this discontinuous dust distribution, with a significant contribution from nano-grains inside 10~au and from usual 'larger' grains outside 10~au? 


Dust filtering processes have been often invoked to explain spatial discrimination between different grain sizes in discs \citep[e.g.,][]{2006Rice}. In the general scheme, large grains, i.e. partially decoupled from the gas, are trapped in gas pressure bumps. The latter can have different origins such as: the presence of a planet that creates a strong gas pressure/density gradient at the inner edge of the gap/cavity induced by the planet \citep[e.g.,][]{2015Pinilla}; magnetohydrodynamic instabilities like the Rossby-wave instability \citep[e.g.,][]{2010Meheut}, triggered by strong gas density gradients and inducing pressure bumps (anticyclonic vortices) that can trap preferentially large dust grains; or self-induced dust traps \citep{2017Gonzalez}.
Determining which process is at the origin of the apparent dust filtering in the HD\,179218 disk is beyond the scope of this work and will require additional data. In particular, the perspective of N-band closure phases revealing possible brightness asymmetries at the inner edge of the outer disk would help disentangle the different processes. 


Our model supports the presence of such nano-grains in an optically thin region strongly irradiated by a very luminous young star. Such very small grains are a priori easily destroyed by UV photons \citep[e.g.,][]{2001Munoz, 2001Menella, 2012Gadallah, 2014Alata} or blown away by the stellar radiation pressure \citep[e.g.,][]{2002Koehler}. Hereafter we provide possible explanations for their presence.



\begin{figure*}
\sidecaption
  \includegraphics[width=12cm]{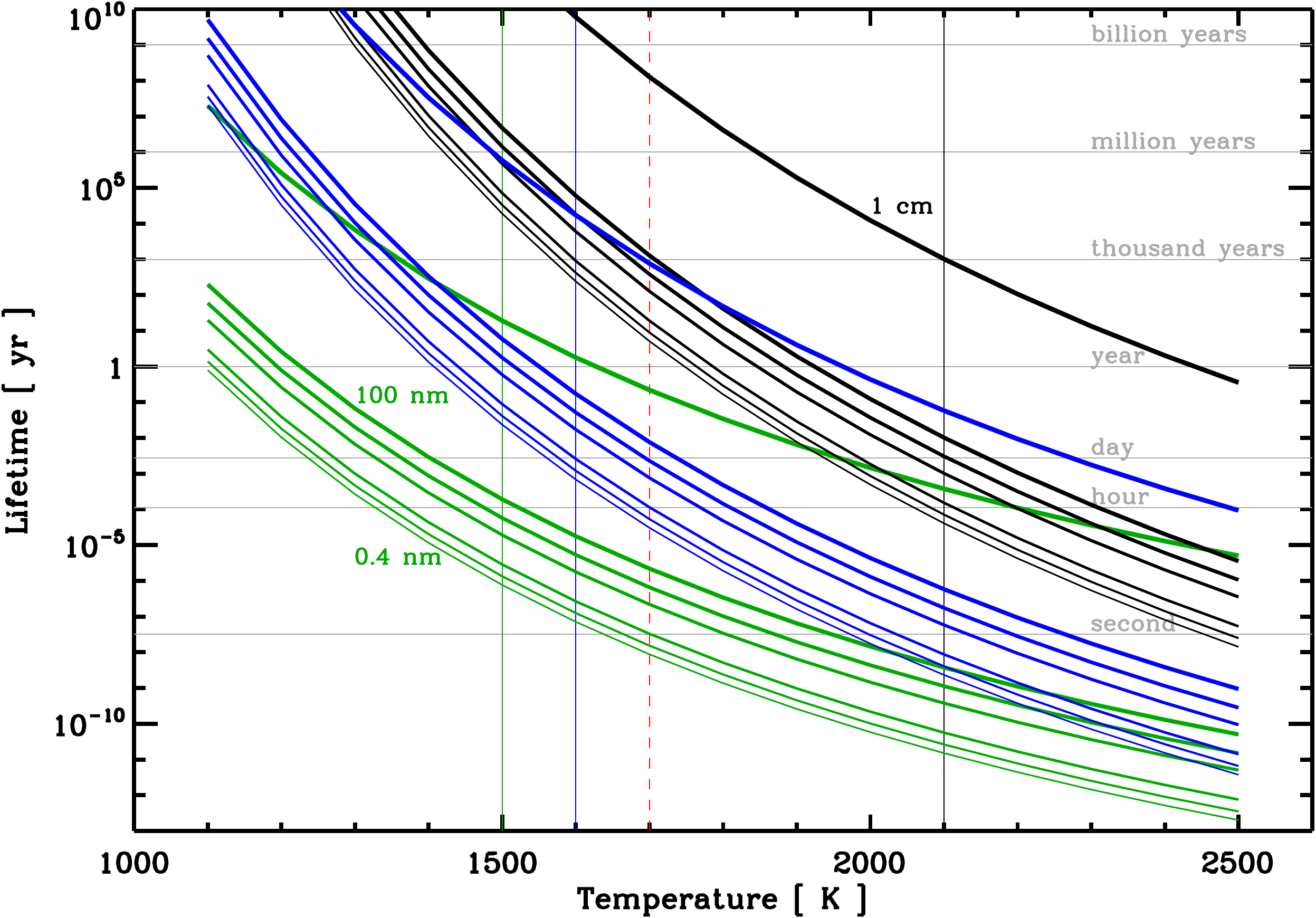}
     \caption{Calculated lifetimes for silicates (green), graphite (black) and THEMIS a-C (blue).
The thicker the line the bigger the grains:  0.4, 0.7, 1.5, 10, 30, 100~nm and the thickest for 1cm radius
particles. The solid vertical lines represent the “classical” sublimation temperatures 
of the three dust materials. The red dashed line is the apparent “model required” temperature of 1700~K. Refer to \citet{2009Kobayashi} to see the calculation of the lifetimes.}
     \label{fig:themis_lifetime}
\end{figure*}

A possibility is that mostly small grains (probably sub-$\mathrm{\mu m}$-sized and smaller) are passing through the 10~au 'barrier' and are flowing inward. Fig.~\ref{fig:themis_Tdust} shows that the smallest carbonaceous nano-grains (< 1~nm) are expected to reach their sublimation temperature right after passing the 10~au barrier, their survival lifetime is around one year (see Fig.~\ref{fig:themis_lifetime}).
At such distance, the 10~nm grains are much colder and their survival lifetime is longer than $10^6$~years. The IR emission at 10~au would thus be dominated by the smallest, and thus hottest, grains close to their sublimation temperature. Then with the inward accretion flow, such grains would sublimate quickly and the 'new' smallest grains present in the population would replace them and so forth. That would leave, at each radius, only larger grains that are still cold enough to survive and keep drifting inward. Ultimately, the largest grains of the initial population would sublimate at the closest distance to the star. 

Among the different accretion rate estimates of HD\,179218, the most recent one gives $10^{-6}$~$\mathrm{M_{\odot}}$/yr \citep{2020Wichittanakom}. Assuming that such an accretion rate is closely associated with the accretion flow from the outer disk regions, $10^{-6}$~$\mathrm{M_{\odot}}$ of gas would be provided to the inner 10~au region per year. That translates to $10^{-8}$~$\mathrm{M_{\odot}}$ of dust per year, assuming the typical diffuse ISM dust/gas ratio of 1/100. 
Given the inner disk dust mass of $M_1= 3.1 \cdot 10^{-10} ~\mathrm{M_{\odot}}$ we derived from our modeling (see Section~\ref{sec:sensitivity}) and the estimated grain lifetimes mentioned above, the accretion flow from the outer regions could provide a significant amount of the necessary dust grains population. We note that spatial variations of the dust/gas ratio are expected in discs. Values lower than $10^{-2}$ are predicted for the outer disk regions ($\gtrsim 100$~au), as a result of efficient dust radial drift \citep[e.g.,][]{2012Birnstiel,2012Hugues}, or derived from mm observations \citep{2019Powell}. However, ISM-like or even enhanced dust/gas ratios are expected inward due to efficient grain fragmentation \citep[e.g.,][]{2012Hugues}. Such a trend remains consistent with our conclusion on the inner disk replenishment in dust.




Such continuous replenishment of dust from disk accretion is likely combined with other local processes like the regeneration of nm-sized grains by the partial photo-fragmentation of larger aggregates \citep{2020Schirmer}. Photo-fragmentation is expected to liberate, from one larger grain (e.g., mm-sized), billions of hot nano-particles that will thus dominate the IR emission. 


\subsection{AU-scale location and characterization of the nano-grains}


 According to Fig.~\ref{fig:themis_lifetime}, the lifetime of silicate nano-grains is expected to be much shorter than THEMIS grains or even pure graphite grains. That would strongly prevent their survival in the inner 10~au region and leave a population of carbon-rich nano-grains. Interestingly, the MIDI correlated fluxes of HD\,179218 show a prominent silicate feature around 10 micron at the smallest baselines, which seems then to fade out at the longer ones \citep{2018Varga}. This supports the fact that the inner 10~au region is probably silicate-depleted with more refractory dust species being favoured like carbon or iron.
 

 To characterize the composition and the properties of the species comprising the nano-grain population, together with their location in the inner disc, we can thus focus on spectral features from the carbonaceous material, especially the aromatic and aliphatic features. In such a way, determining whether the 3.3~$\mathrm{\mu m}$ and 3.4~$\mathrm{\mu m}$ features have different spatial distributions, with respect to the continuum emission, will indicate the spatial segregation of carbonaceous compounds as we approach the star. In that context, a key question is: does hydrogenation vary with
radial distance to the star?
 
In THEMIS, the nano-carbons responsible for the aromatic and aliphatic bands are one and the same population of which the hydrogen content can vary with size and/or distance from the star. They all emit continuum in the NIR. By going from highly aromatic grains (they therefore resemble PAHs) which are suitable for the diffuse medium ($E_g = 0.1$ eV $\Leftrightarrow X_H = 0.02$) to much more hydrogenated grains, capable of emitting strongly at 3.4 $\mathrm{\mu m}$ ($E_g = 0.5$ eV $\Leftrightarrow X_H = 0.12$), this continuum emission decreases by a factor of 10.
 
\citet{2019Bouteraon} found that aliphatic bonds, which are more fragile than the aromatic ones, are expected to break under the strong UV irradiation like in the inner gap of HD\,179218. However, we find no significant variations in the aliphatic-to-aromatic emission band ratios from their large scale L-band spectra observations. This suggests that there should be a continuous replenishing mechanism which could be due, in part, to the fragmentation of larger grains at the surface of the disk at larger scales (see Section \ref{sec:two grains}).

In Section \ref{sec: PAHs}, we could provide first constraints on the presence and location of aromatic grains (based on the 3.3~$\mathrm{\mu m}$ feature) from the existing low resolution MATISSE data. Assuming a global $r^{-2}$ intensity profile for the aromatic band, the existing L-band data in LR are indeed compatible with the existence of aromatic grains provided such grains are located farther than 1~au from the central star.
As mentioned previously, depending on the distance of carbonaceous nano-particles to the star we might expect the aliphatic-to-aromatic band ratio to change which reflects the processing of the hydrocarbon grains a-C:H by UV photons (\citet{2019Bouteraon}, \citet{2020Habart}).
Therefore, from simulated medium-resolution MATISSE data, we assessed the feasibility of a more accurate characterization (location, band ratio) of the emitting regions associated with the aromatic component at 3.3~$\mathrm{\mu m}$ and to the aliphatic one at 3.4~$\mathrm{\mu m}$. Assuming a 30$\%$ emission contribution from the carbonaceous emission bands (with respect to the IR continuum), as detected by ISO/SWS \citep{2006Acke}, we would be able to discriminate between different inner radii of emission down to the au scale. 
Moreover, medium resolution observations with MATISSE will allow to discriminate better between the aromatic and aliphatic band signatures, as illustrated in Fig.~\ref{fig:scaled PAH in MR}, and will allow to determine the location of their emission. With a determination, at the au-scale, of their location and the radial evolution of their band ratio, strong constraints are expected on the nature of the associated carbon nano-grains (e.g. size, hydrogenation state, structure) as a function of the distance from the star. Probing those grains at such close distances from the star and such small scale is a unique opportunity to study the photo-induced fragmentation of large hydrogenated amorphous carbon grains (containing aliphatic structures) along with the dehydrogenation and destruction processes of carbon grains. Notably, identifying clearly where the C-H induced features start in HD\,179218's disk will be key. Interestingly, a complete nondetection of those features in the future MATISSE data would point toward very efficient dehydrogenation processes and/or the presence of another dominant refractory species like iron.
What is at stake is a better understanding of the physical conditions under which solid carbon can survive in the terrestrial planet-forming regions of discs. 





Future observations with MATISSE at higher spectral resolution will thus shed light on the circumstellar environment of HD\,179218.
Additional near-infrared interferometric constraints coupled with a self-consistent radiative transfer modeling with POLARIS \citep{2017ABrauer} will give clear answers on the existence of nano-carbons in the inner component. 
POLARIS includes the consideration of complex structures and varying dust grain properties in circumstellar discs as well as the stochastic heating of nm-sized dust grains \citep{2019Brauer_disk}. 

	\section{Summary and conclusion}
	\label{Conclusions}
	We presented the first L-band interferometric observations of the circumstellar emission around HD\,179218 using the VLTI instrument MATISSE. Our results are the following:
	\begin{itemize}
\item A two-component disk model, featuring temperature and surface density radial profiles, could consistently reproduce both the broadband SED (visible and IR) and H-band (PIONIER), L-band (MATISSE), and N-band (MIDI) visibilities.
\item From our temperature-gradient modeling, we confirmed that the H-band and L-band emitting region is spatially extended ($\sim$ 10~au) and has a homogeneously high temperature ($\sim$ 1700~K). We also confirmed the ring-like shape observed in the N-band continuum emitting region, which starts at about 10~au and with a temperature of $\sim 400$~K, expected for usual $\mathrm{\mu m}$-sized grains located at such distance.
\item Our results suggest the presence of two separate dust populations: a region filled with stochastically heated very small (nano) carbon grains within 10~au can maintain a very high temperature.; and a colder outer disk for which the IR emission is dominated by larger, probably $\mathrm{\mu m}$-sized, grains. 
\item With the previous detection of carbonaceous nano-grains down to 30~au in HD\,179218's disc, their survival in the highly irradiated inner 10~au region brings new insights in the different photo-induced processes acting on such small grains, and their regeneration/replenishment. Considering the expected lifetimes of the THEMIS carbonaceous grains at the temperatures reached in the inner 10~au region, the HD\,179218's accretion rate could maintain a significant inflow of material to maintain a stable nano-grain population there. Moreover, local photo-fragmentation of larger carbonaceous grains could play a significant role in the regeneration of nano-grains. 
\item First constraints could be provided on the presence and location of aromatic grains 
from the existing low resolution MATISSE data. Indeed, we found that the MATISSE data are not compatible with the presence of aromatic grains in close proximity to the star (closer than about 1 au). 
	\end{itemize}
	
The perspective of such a study is a better understanding of the physical conditions under which solid carbon can survive in the terrestrial planet-forming regions. To this end, we simulated MATISSE L-band visibilities in medium spectral resolution to assess the feasibility of detection and detailed characterization of the aromatic (3.3~$\mu$m) and aliphatic (3.4~$\mu$m) band emitting regions in the inner 10 au. We expect at the au-scale their location and the radial dependence on their band ratio. Strong constraints are thus expected on the nature of the associated nano-carbon grains (size, hydrogenation state, structure) as a function of the distance from the star. Especially, identifying clearly where the C-H induced features start in HD\,179218's disk will inform us on the dehydrogenation and destruction processes of carbon grains at close distances from the star.

\begin{acknowledgements}
MATISSE is defined, funded and built in close collaboration with ESO, by a consortium composed of French (INSU-CNRS in Paris and OCA in Nice), German (MPIA, MPIfR and University of Kiel), Dutch (NOVA and University of Leiden), and Austrian (University of Vienna) institutes. The Conseil Départemental des Alpes-Maritimes in France, the Konkoly Observatory and Cologne University have also provided ressources for the manufacture of the instrument. This work has been supported by the French government through the
UCAJEDI Investments in the Future project managed by the National research Agency (ANR) with the reference number ANR-15-IDEX-01. We acknowledge Université Côte d'Azur for the PhD funding of E.Kokoulina via the program "Center For Planetary Origin" (C4PO).
E. Habart, A.P. Jones and N. Ysard acknowledge the support of the Programme National PCMI of CNRS/INSU with INC/INP co-funded by CEA and CNES.
P. \'Abrah\'am acknowledges the support of the Hungarian NKFIH grant  K132406.
A thought goes to our two nice colleagues, Olivier Chesneau and Michel Dugué, at the origin of the MATISSE project with several of us and with whom we shared plenty of beautiful moments.
\end{acknowledgements}

 \institute{
Laboratoire Lagrange, Universit\'e C\^ote d'Azur, Observatoire de la C\^ote d'Azur, CNRS, Boulevard de l'Observatoire, CS 34229, 06304 Nice Cedex 4, France\label{inst_O} 
\and
AIM, CEA, CNRS, Universit\'e Paris-Saclay, Universit\'e Paris Diderot, Sorbonne Paris Cit\'e, F-91191 Gif-sur-Yvette, France\label{inst_Pa}
\and
Max-Planck-Institut f\"ur Radioastronomie, Auf dem H\"ugel 69, D-53121 Bonn, Germany\label{inst_B} 
\and Leiden Observatory, Leiden University, Niels Bohrweg 2, NL-2333 CA Leiden, the Netherlands\label{inst_L}
\and
Konkoly Observatory, Research Centre for Astronomy and Earth Sciences, E\"otv\"os Lor\'and Research Network (ELKH), Konkoly-Thege Mikl\'os \'ut 15-17, H-1121 Budapest, Hungary\label{inst_K} 
\and
Max Planck Institute for Astronomy, K\"onigstuhl 17, D-69117 Heidelberg, Germany\label{inst_H} 
\and
Univ. Grenoble Alpes, CNRS, IPAG, 38000, Grenoble, France\label{inst_I} 
\and
Anton Pannekoek Institute for Astronomy, University of Amsterdam, Science Park 904, 1090 GE Amsterdam, The Netherlands\label{inst_P}
\and
Institute for Mathematics, Astrophysics and Particle Physics, Radboud University, P.O. Box 9010, MC 62 NL-6500 GL Nijmegen, the Netherlands\label{inst_Ra} 
\and
NOVA Optical IR Instrumentation Group at ASTRON (Netherlands)\label{inst_A} 
\and
European Southern Observatory, Alonso de Cordova 3107, Vitacura, Santiago, Chile\label{inst_E}
\and European Southern Observatory Headquarters, Karl-Schwarzschild-Stra\ss e 2, 85748 Garching bei M\"unchen, Germany \label{inst_Garch}
\and
SRON Netherlands Institute for Space Research, Sorbonnelaan 2, NL-3584 CA Utrecht, the Netherlands\label{inst_U} \and
Institute of Theoretical Physics and Astrophysics
University of Kiel, 24118 Kiel, Germany
\label{inst_Ki}
\and
NASA Goddard Space Flight Center
Astrophysics Science Division, Code 660
Greenbelt, MD 20771 \label{inst_Nas} 
\and
 IMPMC, CNRS-MNHN-Sorbonne Universit\'es, UMR7590, 57 rue Cuvier, 75005 Paris, France \label{inst_IMPMC}
 \and
Department of Astrophysics, University of Vienna, Türkenschanzstrasse 17, A-1180 Vienna, Austria\label{inst_V}
\and
I. Physikalisches Institut, Universit\"at zu K\"oln, Z\"ulpicher Str. 77, 50937, K\"oln, Germany\label{inst_C} \and
Institut f\"ur Kernphysik, Universit\"at zu K\"oln, D-50937 K\"oln, Germany\label{inst_IKP} \and
Sydney Institute for Astronomy, School of Physics, A28, The University of Sydney, NSW 2006, Australia\label{inst_Sy}\and
Institut d'Astrophysique Spatiale, CNRS, Univ. Paris-Sud, Universit\'e Paris-Saclay, B\^{a}t. 121, 91405 Orsay cedex, France\label{inst_Saclay}
\and
Institut des Sciences Mol\'eculaires d'Orsay (ISMO), UMR8214, CNRS - Universit\'e de Paris-Sud, Universit\'e Paris-Saclay, Bat 520,
Rue Andr\'e Rivi\'ere, 91405 Orsay, France\label{inst_Mol}
}

	\bibliographystyle{aa} 
	\bibliography{ref} 

\begin{appendix} \label{appendix}
 \section{Parameter sensitivity} \label{sec:sensitivity}
From the global best-fit solution we found and described in Section \ref{subsubsection:results}, here we explore four previously fixed fundamental parameters of the inner component, $q_1$, $p_1$, $R_{in, ~ 1}$ and $p_2$. The reason is twofold : 1) testing the sensitivity of the model to those parameters and thus the robustness of our initial assumptions about the inner component (isothermal aspect, achromatic and radially independent emissivity, expected silicate dust sublimation radius setting the inner radius); 2) improving possibly the agreement with the low-frequency MATISSE visibilities. 
For that, we considered the inner component to be a classical temperature-gradient disk with $q_1$, $p_1$, $M_1$ and $R_{in, ~ 1}$ set as free parameters, and with the same dust composition as for the outer component.

Overall, with the uniform priors for $q_1$, $p_1$ and $p_2$ ranging from 0 to 1 and from 0 to 1.5 (for $p_1$ and $p_2$), respectively, the reduced $\chi^2$ of the best-fit model appears to be worse than in the isothermal case. 
The best-fit value for $p_2$ is $\sim$ 1.3 which is not far from the fixed value we used for the temperature-gradient model. The best-fit value of $q_1$ clearly converged to zero, which validates our initial isothermal assumption for the inner component. However, it is less clear for $p_1$, which shows a flat posterior probability distribution and thus can not be very well constrained by our data. Nevertheless, the derived mass of the inner disc, $M_1= 3.1 \cdot 10^{-10} ~\mathrm{M_{\odot}}$, translates again to a low vertical optical depth at all IR wavelengths. 
That reinforces the optically thin status of the inner 10~au region. The optical depth estimated in the radial direction of the disk (in the disk mid-plane assuming a scale height factor $H_p$ distributing vertically the material) is very low, i.e. the inner component is optically thin (optical depth of about $10^{-3}$ integrated from 2.25 to 7~au). 
Concerning the inner radius $R_{in, ~ 1}$, it was scanned within 0.35 to 4~au. Our analysis of the posterior distribution shows that we can not obtain 
a clear constraint on the inner radius. The 1-$\sigma$ region is rather spread and all the solutions have similar reduced $\chi^2$. We can just give the upper limit on the inner radius: $R_\mathrm{in, ~1} < 3.24$ au. Like that we can find many similar solutions within this 1-$\sigma$ region. It is clear that our L-band MATISSE data could provide constraints on the extension of the inner disk component but remains insufficient to constrain the inner radius of the IR-continuum emission.
The exploration of different values of $R_{in, ~ 1}$ does not provide a better agreement with the low-frequency MATISSE visibilities. The slope of the model, not necessarily well represented by the $\chi^2$ criterium, could be however improved compared to the one of the low frequency data by favoring the greater value of $R_\mathrm{in, ~1}$. But we could not anyway obtain a better adjustement in the context of our present model assumptions concerning its physics and the considered disk geometry.

\section{Closure phase modeling} \label{sec:closure_phase}

\begin{figure*}
 \centering
 \includegraphics[width=\textwidth]{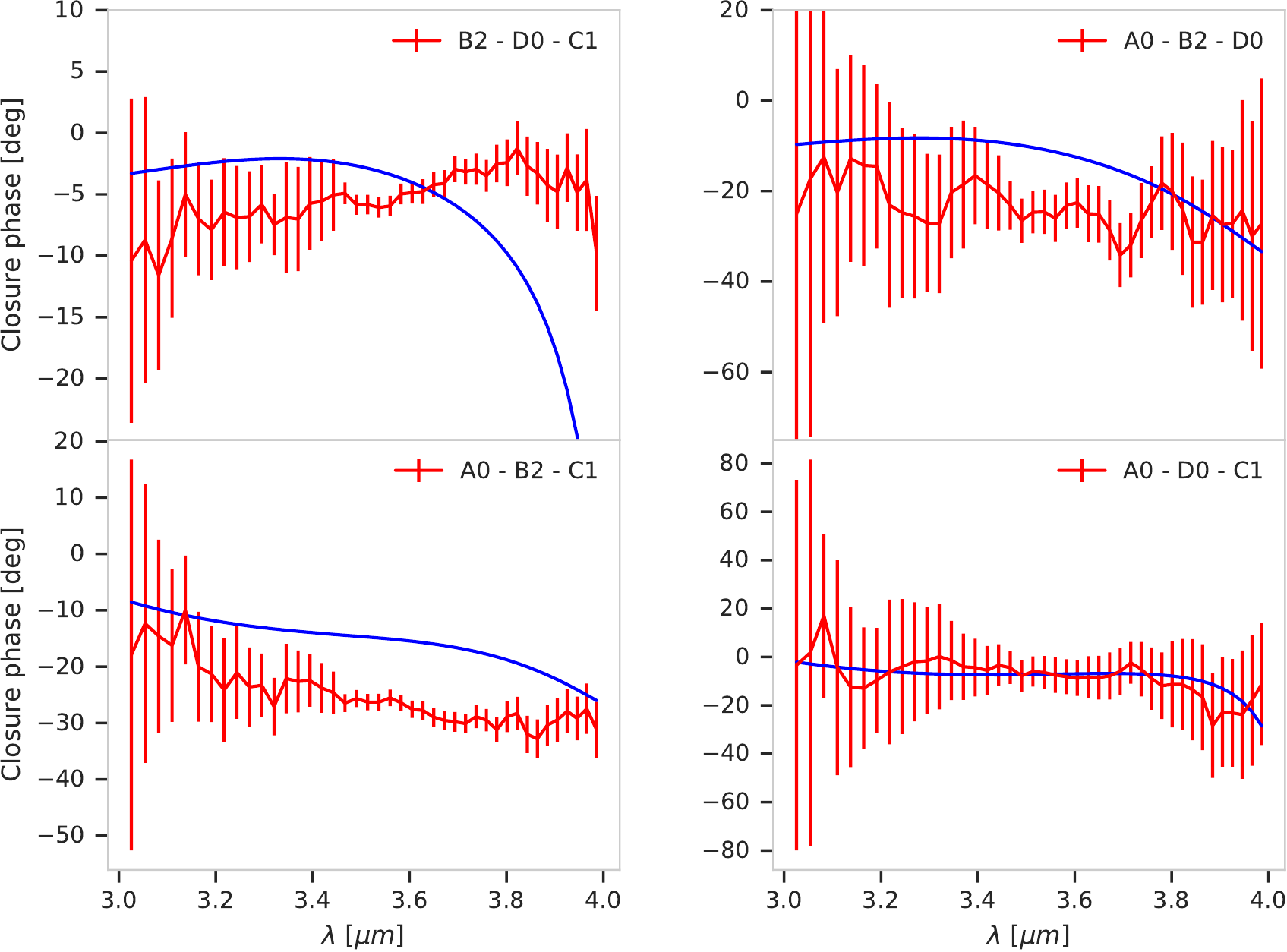}
 \caption{MATISSE closure phases in L-band. Red curve represents the data and the blue line is the fit. The telescopes stations are shown in each of the subplots.}
 \label{fig:closure phase}
\end{figure*}

The closure phase measurements presented in Fig.~\ref{fig:closure phase} in red display an amplitude of several tens of degrees. This is the signature of an asymmetry.

Several hypotheses can be considered in the frame of our centro-symmetrical temperature-gradient model, we tested the possibility that the closure phase signal is due to an apparent shift of the photocenter of the disk brightness by respect to the star. 
We decided to introduce this shift through $\alpha$ and $\beta$, the star position by respect to the disc's central position. 
In this case, the complex and normalized visibility of the star ($V_{star}$) is expressed by:
\begin{equation}
 V_{star} = \exp{\left( -2 i\pi (B_{u, \theta}~ \alpha +B_{v, \theta} ~\beta ) \right) } 
\end{equation}
where $\alpha$ and $\beta$ are the coordinates of the star in the sky in radians. The closure phase is defined as:
\begin{equation}
\psi_{B_1, B_2, B_3} = Arg \left( F_\mathrm{corr, B_1} \cdot F_\mathrm{corr, B_2} \cdot F_\mathrm{corr, B_3}^{\ast} \right)
\end{equation}

where $F_\mathrm{corr, B_1}$ is the sum of the complex correlated fluxes of all the components of the object for the first baseline. 

To fit the closure phase, we used the same approach as the one mentioned in Section \ref{subsec:Fitting approach}. We kept the parameters from the best-fit temperature-gradient model as fixed parameters, and just the $\alpha$ and $\beta$ were left as free parameters. 
We defined the amplitude of uniform priors for $\alpha$ and $\beta$ to be in the range: 0--0.35~au, and after we let those to be much larger than that. The best-fit with the smallest reduced $\chi ^2$ (8.4) (see Fig.~\ref{fig:closure phase}) corresponds to a shift amplitude
larger than 0.35~au (respectively 0.74 and -0.48~au for $\alpha$ and $\beta$ ). 


\begin{figure}
  \resizebox{\hsize}{!}
  {\includegraphics{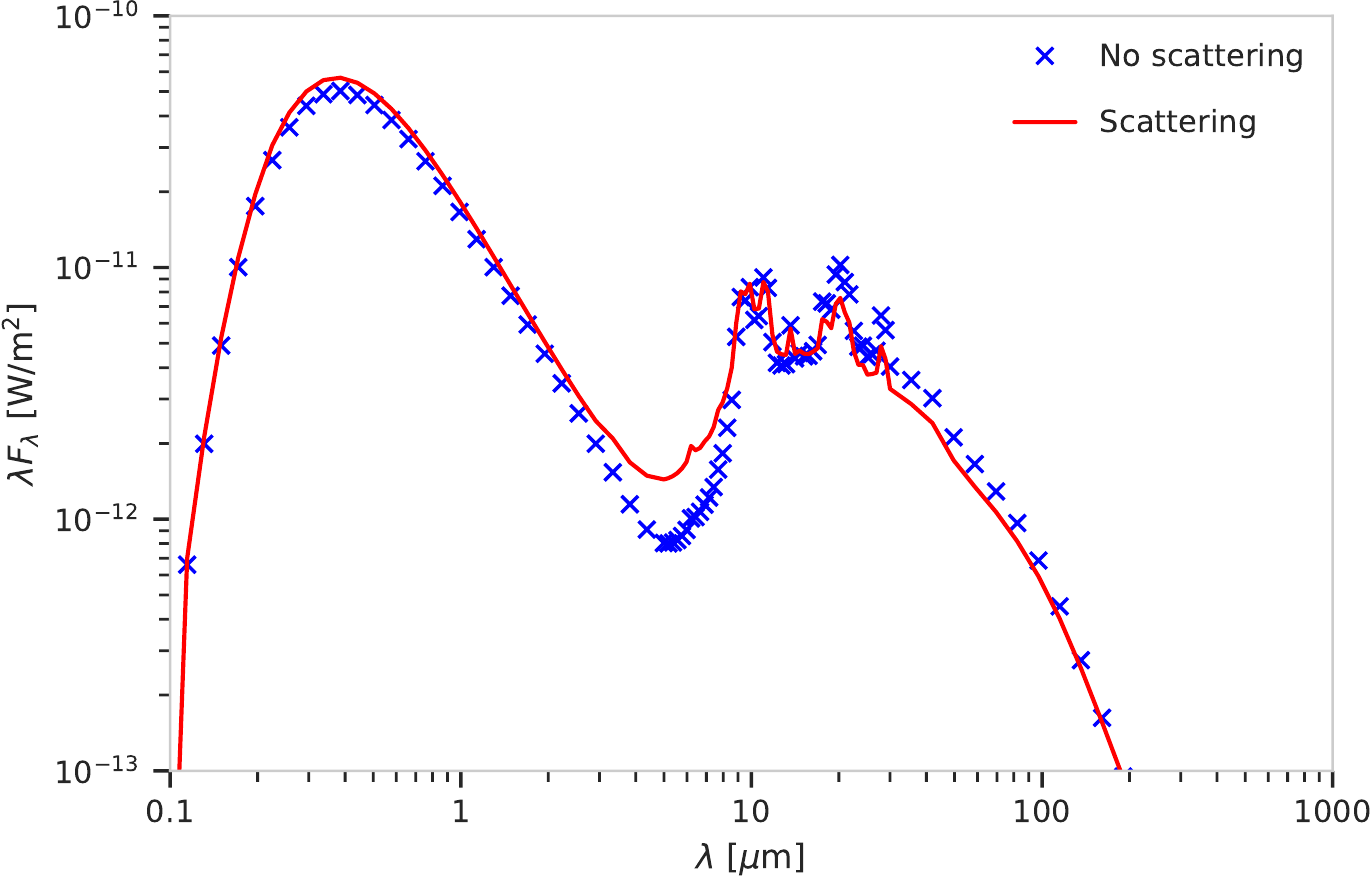}}
  \caption{Radiative transfer simulation illustrating the negligible scattering contribution from very small grains of THEMIS model.}
 \label{fig:RADMC3D}
\end{figure}


\begin{figure}
  \resizebox{\hsize}{!}
  {\includegraphics{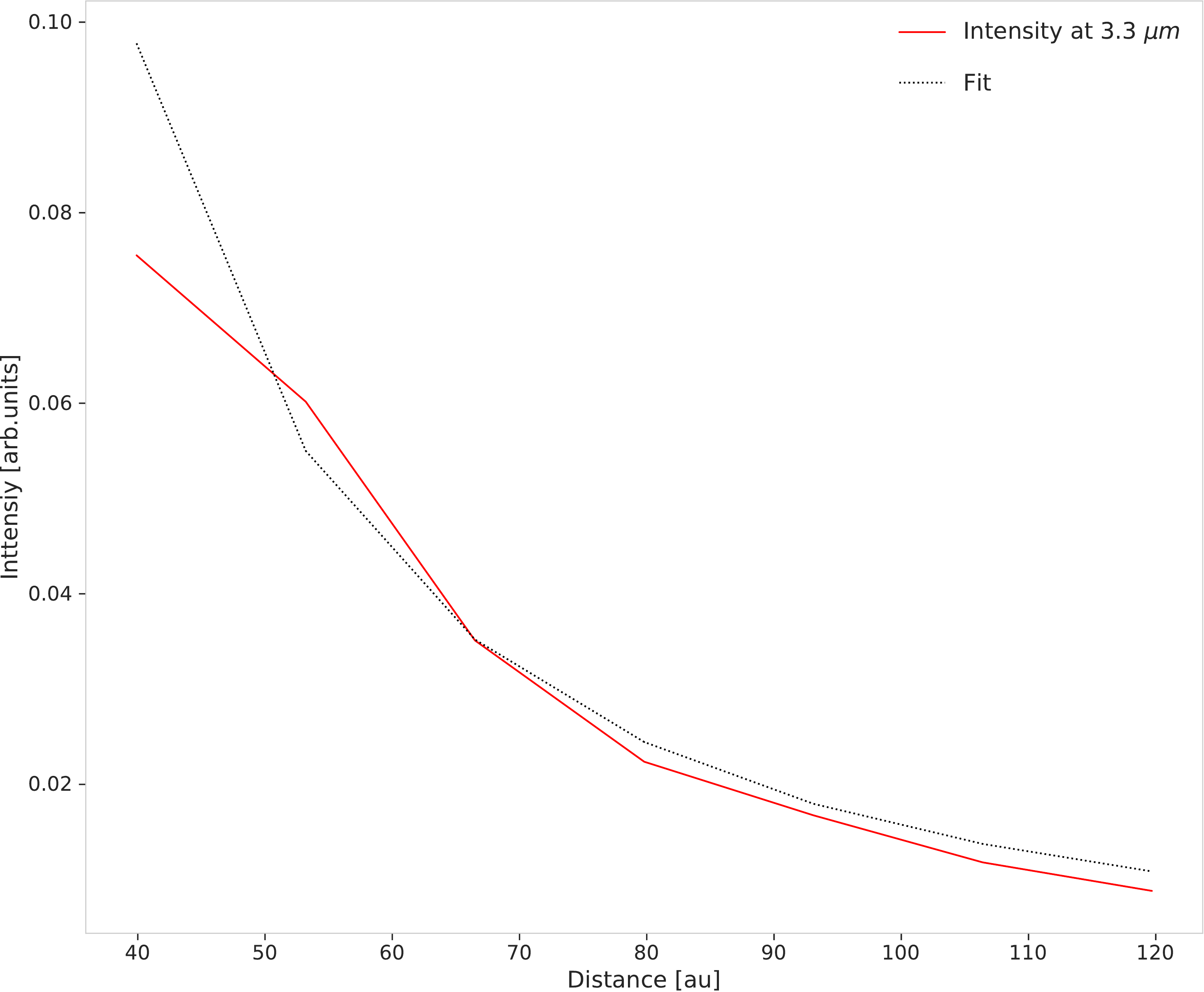}}
  \caption{Intensity profile of HD\,179218 at 3.3~$\mathrm{\mu m}$ vs distance from the star. The red line represents the intensity profile at 3.3~$\mathrm{\mu m}$ calculated from the VLT/NACO data presented in \citet{2019Bouteraon}, and the dotted black line shows the power-law fit in $r^{-2}$.}
 \label{fig:intensity}
\end{figure}

\begin{figure*}
 \centering
 \includegraphics[width=\textwidth]{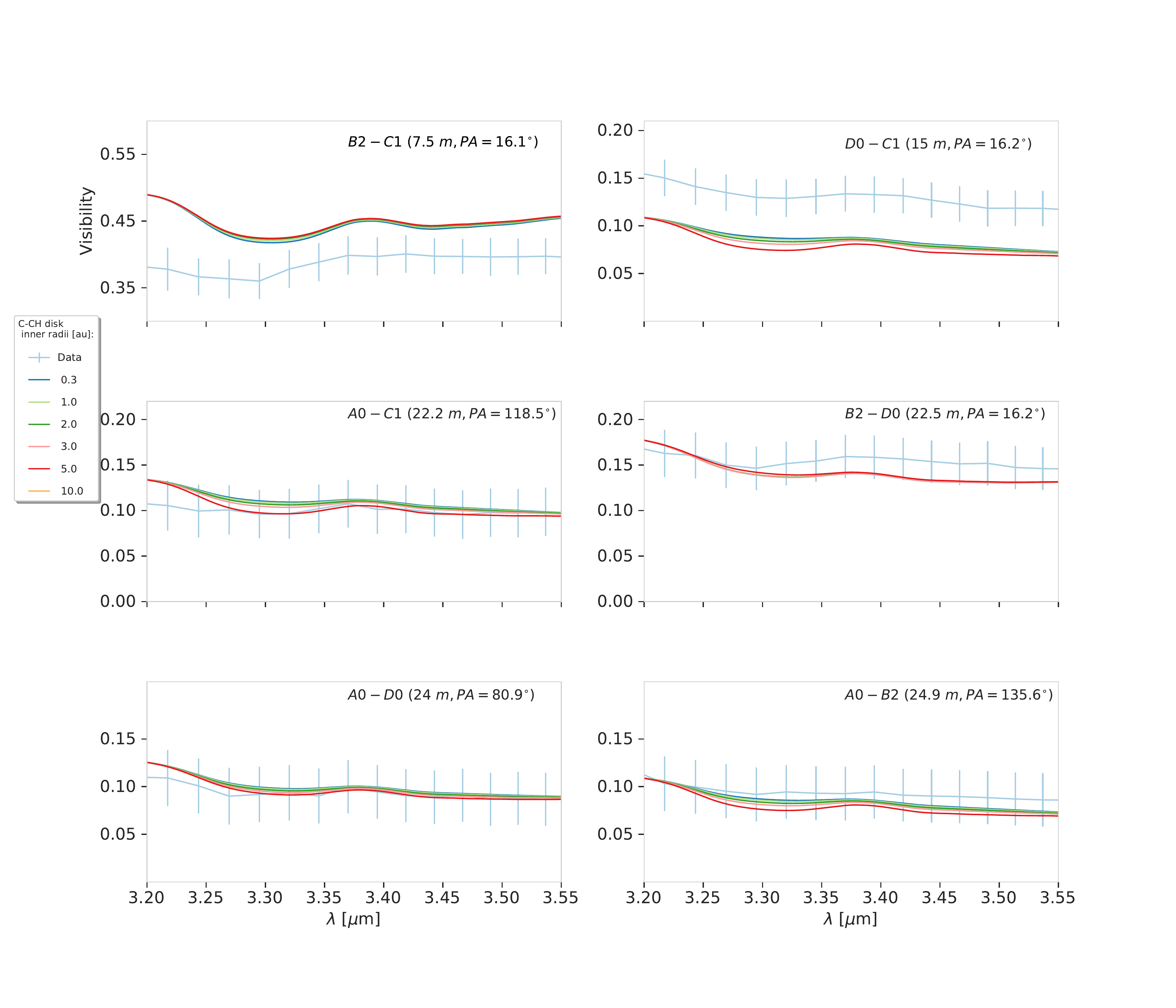}
 \caption{Simulated low resolution ($R$=30) MATISSE visibilities based on the model combining the best-fit continuum model from Section~\ref{subsubsection:results}, with a C-CH disk component from Section~\ref{sec:new intensity prof}. Different inner radii of the C-CH disk with a flat intensity profile are explored - see the legend on the left hand side, the inner radius of the second. The telescope pairs, projected baseline length ($B$) and the position angles of the projected baseline ($B_{PA}$) are shown in each subplot.}
 \label{fig:LR_new_intensity}
\end{figure*}

\section{Data log}\label{app:data log}
\begin{table}\centering
\captionof{table}{Log of observations for the MIDI data}
\begin{tabular}{c|c|c|c}
Instrument & Date [UT] & Telescope's configurations & Filter \\
\hline
MIDI & 2003-06-16 & UT1-UT3 & N\\
  & 2004-04-10 & UT2-UT3 & N\\
	 & 2006-05-15 & UT2-UT3 & N\\
	 & 2006-05-16 & UT1-UT3 & N\\
	 & 2006-05-17 & UT3-UT4 & N\\
	 & 2006-06-11 & UT3-UT4 & N\\
	 & 2006-06-14 & UT1-UT2 & N\\
	 & 2006-07-09 & UT3-UT4 & N\\
 	 & 2006-07-13 & UT1-UT2 & N\\
 	 & 2009-08-14 & E0-G0 & N\\
  & 2009-08-15 & H0-G0 & N\\
  \hline
 \end{tabular}
 \end{table}

\begin{table}
\captionof{table}{Log of observations log for the PIONIER data}
\begin{tabular}{c|c|c|c}
Instrument & Date [UT] & Telescope's configurations & Filter \\
\hline
PIONIER & 2017-04-24 & D0-G2-J3-K0 & H\\
		& 2017-04-24 & D0-G2-J3-K0 & H\\
	 
	& 2017-07-27 & A0-G1-J2-J3 & H\\
	& 2017-07-27 & A0-G1-J2-J3 & H\\
  
	& 2017-09-20 & A0-G1-J2-J3 & H\\
	& 2017-09-20 & A0-G1-J2-J3 & H\\
  \hline
 \end{tabular}
\end{table}

\begin{figure*}
\sidecaption
  \includegraphics[width=12cm]{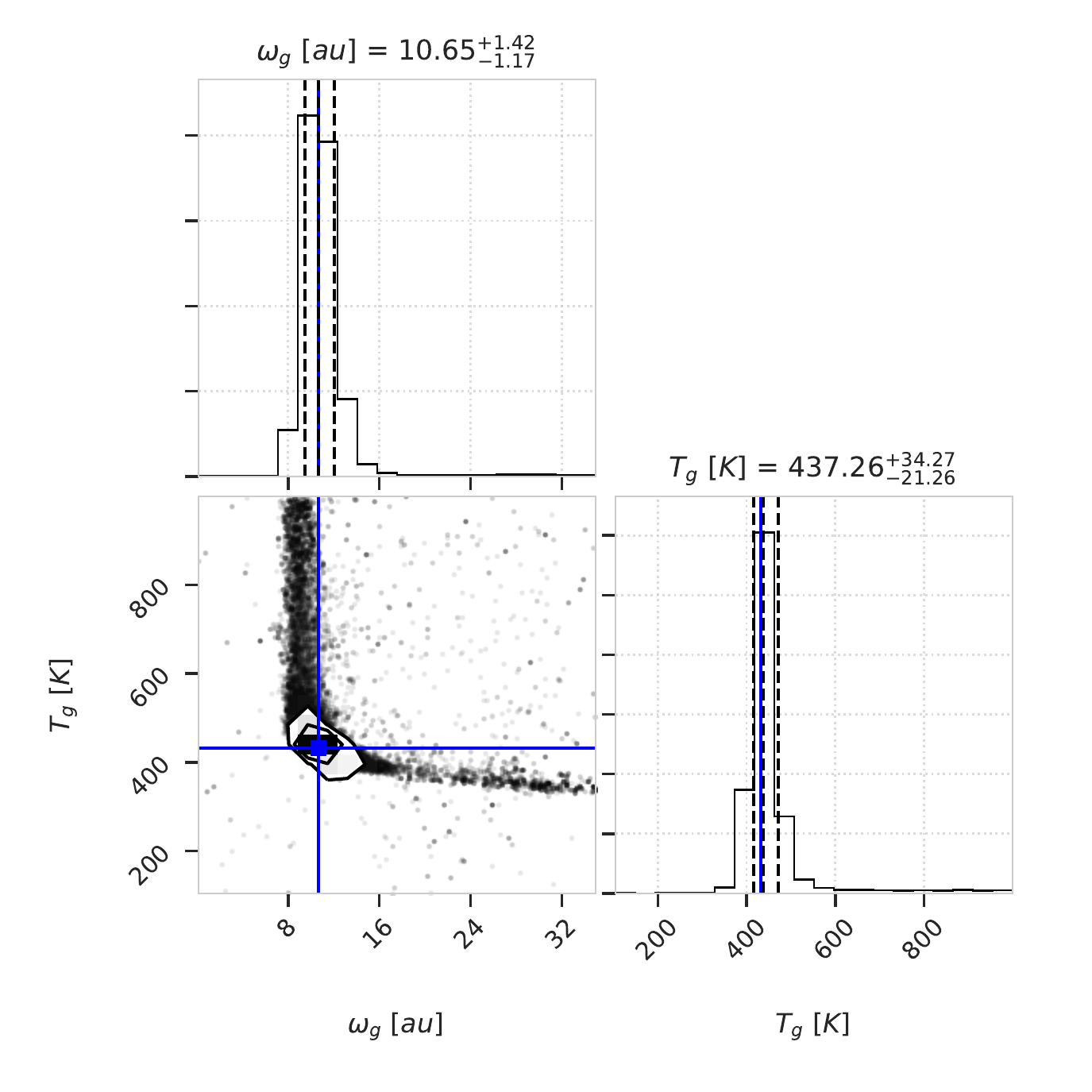}
     \caption{Corner plot showing the one- and two-dimensional posterior distributions for each free parameter in the one-component geometrical model - Gaussian. The run was made with 1000 steps and 100 walkers. The first 200 steps are considered to be "burn-in" phase, and 800 steps are shown where the walkers converged. The blue line shows the best-fit parameters. }
     \label{fig corner:1_comp_gaus}
\end{figure*}


\begin{figure*}
\sidecaption
  \includegraphics[width=12cm]{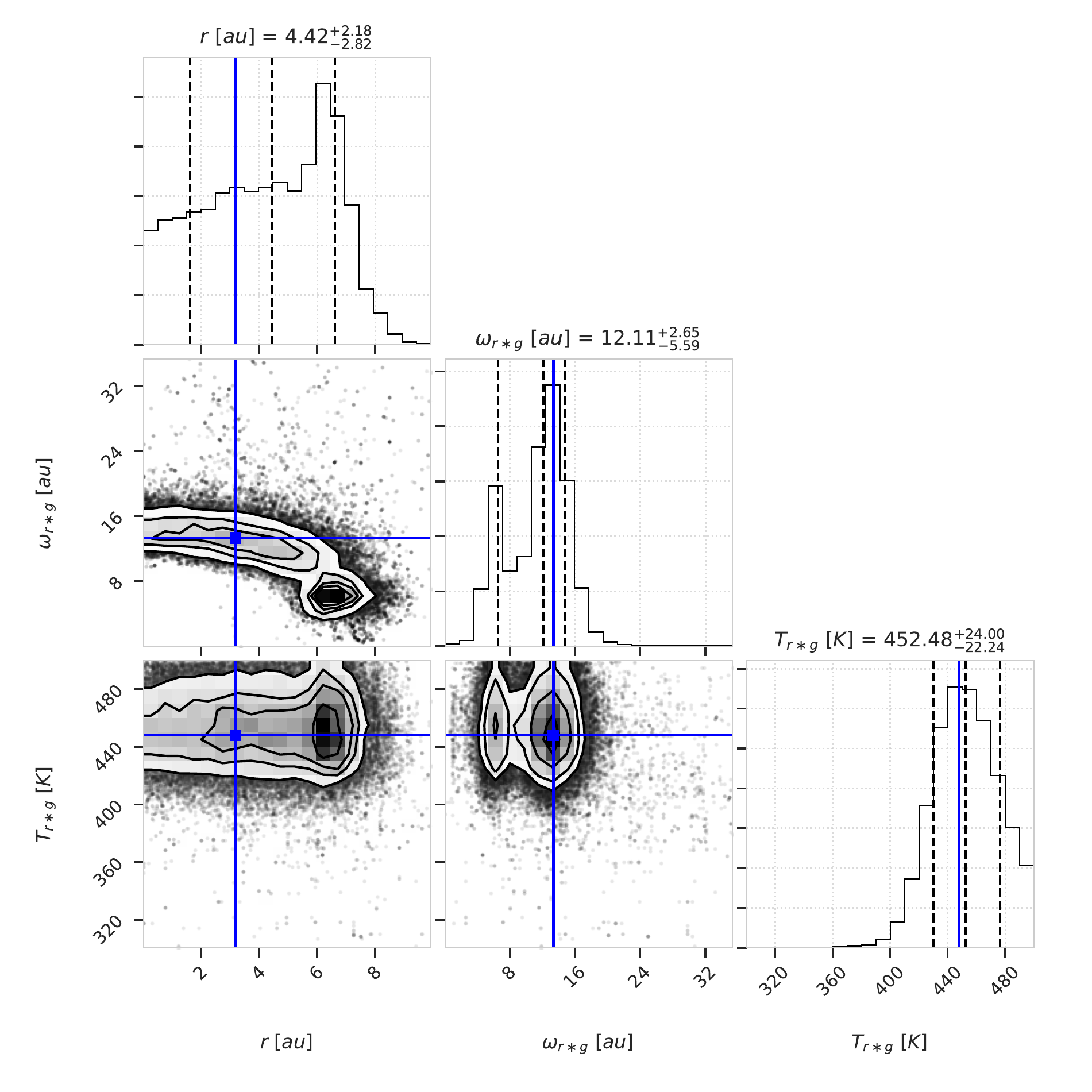}
     \caption{Corner plot showing the one- and two-dimensional posterior distributions for each free parameter in the one-component geometrical model - a ring convolved with a Gaussian. The run was made with 1000 steps and 100 walkers. The first 200 steps are considered to be "burn-in" phase, and 800 steps are shown where the walkers converged. The blue line shows the best-fit parameters. }
     \label{fig corner:1_comp_ring_gaus}
\end{figure*}

\begin{figure*}
 \centering
    \includegraphics[width=17cm]{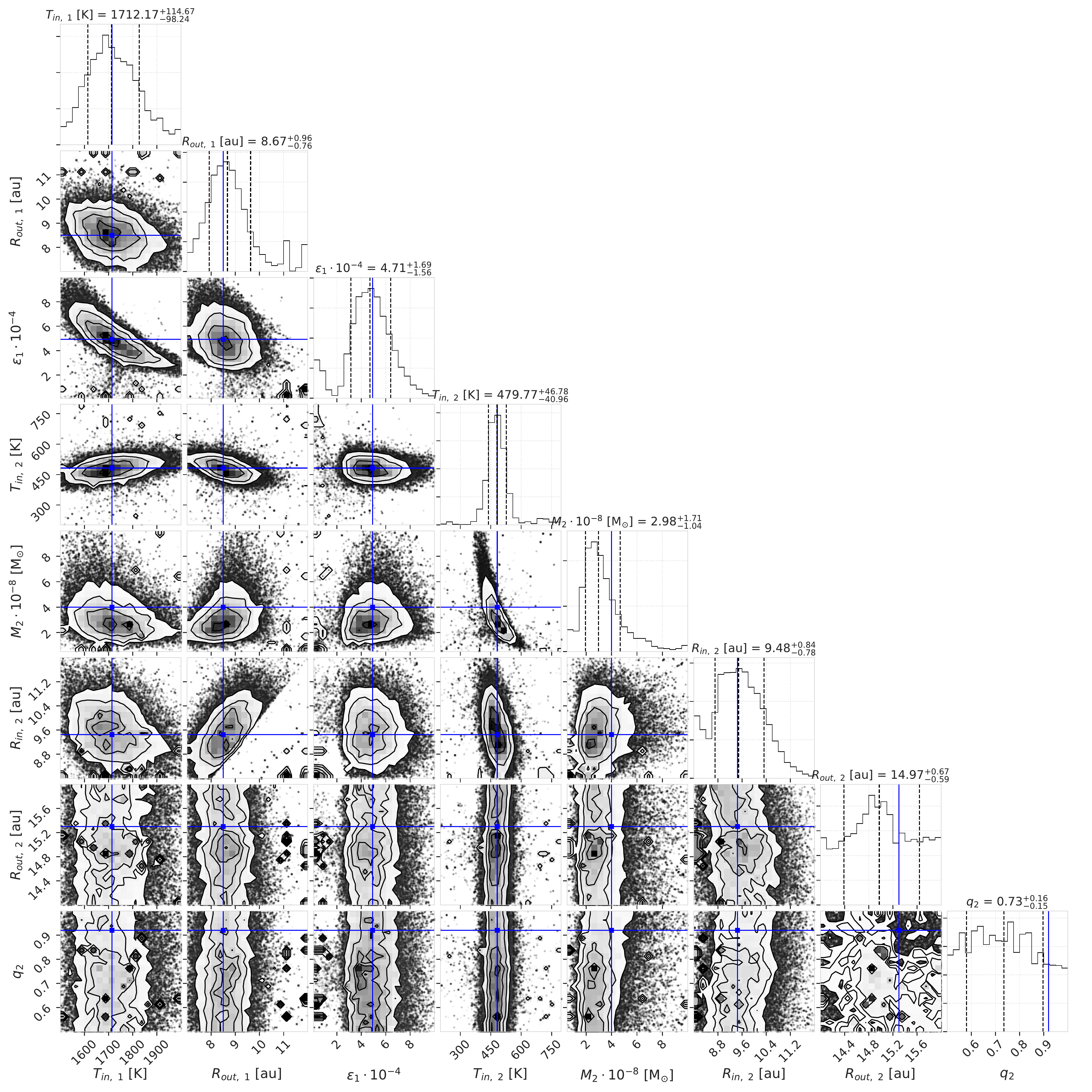}
 \caption{Corner plot showing the one- and two-dimensional posterior distributions for each free parameter in two-component  temperature-gradient model: isothermal inner and a temperature-gradient outer components model for numerous data set: PIONIER + MATISSE + MIDI +SED. The run was made with 2000 steps and 150 walkers. The first 200 steps are considered to be "burn-in" phase, and 1800 steps are shown where the walkers converged. }
 \label{fig:temperature-gradient}
\end{figure*}



\end{appendix}

		
\end{document}